\documentclass[10pt]{iopart}

\usepackage{graphicx}
\usepackage{dcolumn}
\usepackage{amssymb} 

\usepackage{bm}

\newcommand{\cmnt}[1]{}

\def\vecx{\mathbf{x}}
\def\veck{\mathbf{k}}
\def\hatn{\mathbf{\hat n}}
\def\hatk{\mathbf{\hat k}}

\def\VEV#1{\left\langle #1 \right\rangle}
\def\angs{\buildrel _{\circ} \over {\mathrm{A}}}
\def\bs{\!\!\!\!\!\!\!\!\!\!\!\!\!\!\!\!\!\!\!\!\!\!\!\!\!\!\!\!\!\!\!\!\!\!\!\!\!\!\!\!}

\begin{document}

\title[Anisotropic power spectrum]{Non-detection of a statistically anisotropic power spectrum in large-scale structure}

\author{Anthony R. Pullen and Christopher M. Hirata}

\address{Department of Physics, California Institute of Technology, Mail Code 350-17, Pasadena, CA 
91125, USA}

\ead{apullen@caltech.edu}

\begin{abstract}
We search a sample of photometric luminous red galaxies (LRGs) measured by the Sloan Digital Sky Survey (SDSS) for a quadrupolar anisotropy in the primordial power 
spectrum, in which $P(\veck)$ is an isotropic power spectrum $\bar P(k)$ multiplied by a quadrupolar modulation pattern.  We first place limits on the 5 coefficients 
of a general quadrupole anisotropy.  We also consider axisymmetric quadrupoles of the form $P(\veck) = \bar P(k)\{ 1 + g_*[(\hatk\cdot\hatn)^2-\frac13]\}$ where 
$\hatn$ is the axis of the anisotropy.
When we force the symmetry axis $\hatn$ to be in the direction $(l,b)=(94^\circ,26^\circ)$ identified in the recent Groeneboom {\em et~al.} analysis of the cosmic 
microwave background, we find $g_*=0.006\pm0.036$ ($1\sigma$).
With uniform priors on $\hatn$ and $g_*$ we find that $-0.41<g_*<+0.38$ with 95\% probability, with the wide range due mainly to the large uncertainty of 
asymmetries aligned with the Galactic Plane.  
In none of these three analyses do we detect evidence for quadrupolar power anisotropy in large scale structure.
\end{abstract}

\section{Introduction}
\label{S:intro}

Statistical isotropy (SI) is one of the most standard predictions of structure-formation and inflationary models.  In this hypothesis, the density fluctuations in the 
Universe are a realization of a random field whose statistical properties (e.g. power spectra) are invariant under rotations.  When probing density fluctuations using 
the cosmic microwave background (CMB) temperature, SI is generally assumed in the analysis.  However, searches for violations of SI, or statistical anisotropy, are 
now being performed with increasing precision as the amount of CMB data has grown.  These searches are revealing possible evidence for statistical anisotropy in the 
CMB \cite{preferreddirections,Eriksen:2003db,Hansen:2004vq,Eriksen:2007pc}, including a possible detection of quadrupolar anisotropy \cite{Groeneboom:2008fz, G09, 
Bennett:2010jb}.
In response, many have proposed inflationary and dark-energy theories with parameters that quantify departures from SI 
\cite{Chibisov:1989wb,Chibisov:1990bk,Berera:2003tf,Donoghue:2004gu,Buniy:2005qm,Ackerman:2007nb,Gumrukcuoglu:2007bx,ArmendarizPicon:2007nr,Donoghue:2007ze,Pereira:2007yy,Boehmer:2007ut,PhysRevLett.97.131302,Campanelli:2007qn,PhysRevD.80.023521,Shtanov:2010zr}.
The detection of quadrupolar anisotropy in the CMB now appears to be contaminated by systematic effects (possibly beam asymmetry \cite{Hanson:2010gu}, although this 
is debated \cite{G09}); as such it is desirable to constrain quadrupolar anisotropy by other techniques.

One way to quantify statistical anisotropy is to allow the three-dimensional power spectrum of matter density fluctuations $P(\veck)$ to depend on the direction of 
$\veck$.  This is a full description for Gaussian but anisotropic initial perturbations.  This approach was motivated by the inflationary model of 
Ref.~\cite{Ackerman:2007nb}, a model for which Pullen \& Kamionkowski \cite{Pullen:2007tu} 
constructed parameter estimators for CMB analysis.  As usual, if $\delta(\veck)$ is the Fourier amplitude of the fractional matter density perturbation, an 
anisotropic power spectrum is defined via
\begin{eqnarray} \label{E:fluc}
     \VEV{\delta(\veck)\delta^*(\veck')} = 
    (2\pi)^3\delta_{\rm D}(\veck-\veck')P(\veck)\, ,
\end{eqnarray}
where the angle brackets denote an average over all realizations of the random field, and $\delta_{\rm D}$ is a Dirac $\delta$-function; note that we still preserve the 
assumption that different Fourier modes are uncorrelated (statistical {\em homogeneity}).  A direction-dependent $P(\veck)$ can be decomposed via
\begin{eqnarray}   \label{E:powerspectrum}
     P(\veck)= \bar P(k) 
     \left[1+ \sum_{LM} g_{LM}(k) R_{LM}(\hatk) \right],
\end{eqnarray}
where $\bar P(k)$ is the isotropically averaged matter power spectrum, and $R_{LM}(\hatk)$ (with $L\geq2$) are real spherical harmonics.
Physically, $g_{LM}$ gives 
the magnitude of statistical anisotropy on order $L$, with $M$ giving the direction of that anisotropy.  Also, $g_{LM} = 0$ for odd $L$ because a real scalar density 
field always has $\delta(\veck)=\delta^*(-\veck)$ and hence $P(\veck)=P(-\veck)$.

We define the real spherical harmonics $R_{LM}(\veck)$ in terms of the complex spherical harmonics $Y_{LM}(\veck)$ by
\begin{eqnarray} \label{E:realcomplex}
     R_{LM} = \left\{\begin{array}{ll}
     \frac{1}{\sqrt{2}}(Y_{LM}+Y_{LM}^*)&\mbox{if $M>0$} \\
     Y_{L0}&\mbox{if $M=0$} \\
     \frac{(-1)^M}{{\rm i}\sqrt{2}}(Y_{LM}^*-Y_{LM})&\mbox{if $M<0$}\,;
     \end{array} \right.
\end{eqnarray}
these are easily seen to obey the usual orthonormality rules, but have the advantage of making the $g_{LM}$ coefficients real.  The expressions for $L=2$ are given in Appendix \ref{A:realhar}.

The purpose of this paper is to measure or constrain the anisotropy using large scale structure data.  Given the recent debate over the detection of quadrupolar 
anomalies in WMAP, and the evidence that the signal is contaminated by systematic effects \cite{G09, Bennett:2010jb}, it is worth using other datasets as well to 
constrain models with anisotropic power.  In this paper we will assume for simplicity that $g_{LM}$ is 
scale-invariant.  This is both a simplifying assumption, but is also a good first approximation in at least some classes of modified inflationary models 
\cite{Ackerman:2007nb}.  We will also focus only on the quadrupole anisotropy $g_{2M}$; this is the phenomenologically simplest type of anisotropy allowed, and also 
emerges from anisotropic inflation models in the limit of very weak anisotropy \cite{Ackerman:2007nb}.

Galaxy surveys probe matter fluctuations because on large scales, the galaxy density is related to the matter density in accordance with a linear bias model:
\begin{eqnarray} \label{E:bias}
     \VEV{\delta_g(\veck)\delta_g^*(\veck')} = 
     (2\pi)^3\delta_D(\veck-\veck')b_g^2P(\veck)\, ,
\end{eqnarray}
where $\delta_g(\veck)$ is the Fourier amplitude of the fractional galaxy density perturbation, and $b_g$ is the linear galaxy bias.  The galaxy survey probe has been 
used to estimate $P(\veck)$ by stacking the measured angular matter power spectra $C_l$ in eight photometric redshift slices ranging from $z=0.2 \mbox{ to } 0.6$ 
\cite{Padmanabhan:2006ku}.  By performing a similar analysis and including the anisotropy parameters $g_{2M}$ in the power spectrum, we can use galaxy surveys to 
estimate quadrupole anisotropy while assuming fiducial values for the other cosmological parameters.

The plan of our paper is as follows: In Section \ref{S:sample} we describe the SDSS data used and why we choose LRGs to trace the galaxy distribution.  Section \ref{S:form} calculates the angular 
correlations statistical anisotropy produces in galaxy surveys and constructs estimators of the $g_{2M}$s and other systematic power spectrum variations.  We present estimates of these parameters in 
Section \ref{S:result}, and in Section \ref{S:conclude} we present our conclusions.  Wherever not explicitly mentioned, we assume a flat $\Lambda$CDM cosmology with $\Omega_M = 0.3$, $\Omega_b = 0.05$, 
$h = 0.7$, $n_s = 1.0$, and $\sigma_8 = 0.9$.  Since ours is a search for anisotropy, small changes in the cosmology will result only in changes in the calibration 
of the $g_{2M}$ estimator; they do not alter the null hypothesis.

\section{Choice of sample} \label{S:sample}

There are several ways to use galaxy survey data to search for statistical anisotropy.  In principle, one could use a 3-dimensional redshift survey and search for 
anisotropy in the power spectrum.  This would however be very technically involved: redshift-space distortions make the line of sight direction special.  With 
sufficient sky coverage one could break the distinction between redshift-space distortions and true statistical anisotropy.  However, in this paper we choose the 
technically simpler route of using photometric galaxy catalogues, which can be studied using estimators analogous to those for the CMB.

The photometric data we use come from the Sloan Digital Sky Survey (SDSS) \cite{York:2000gk}.  The SDSS consists of a 2.5 m telescope \cite{Gunn:2006tw} with a 
5-filter (\textit{ugriz}) imaging camera \cite{Gunn:1998vh} and a spectrograph.  Automated pipelines are responsible for the astrometric solution \cite{Pier:2002iq} and 
photometric calibration \cite{Fukugita:1996qt, Hogg:2001gc, Tucker:2006dv, Padmanabhan:2007zd}.  Bright galaxies, luminous red galaxies (LRGs), and quasars are 
selected for follow-up spectroscopy \cite{Strauss:2002dj, Eisenstein:2001cq, Richards:2002bb, Blanton:2001yk}.  The data used here were acquired between August 1998 
and October 2004 and are included in SDSS Data Release 5 \cite{AdelmanMcCarthy:2007wh}.

\begin{figure}
{\scalebox{.8}{\includegraphics{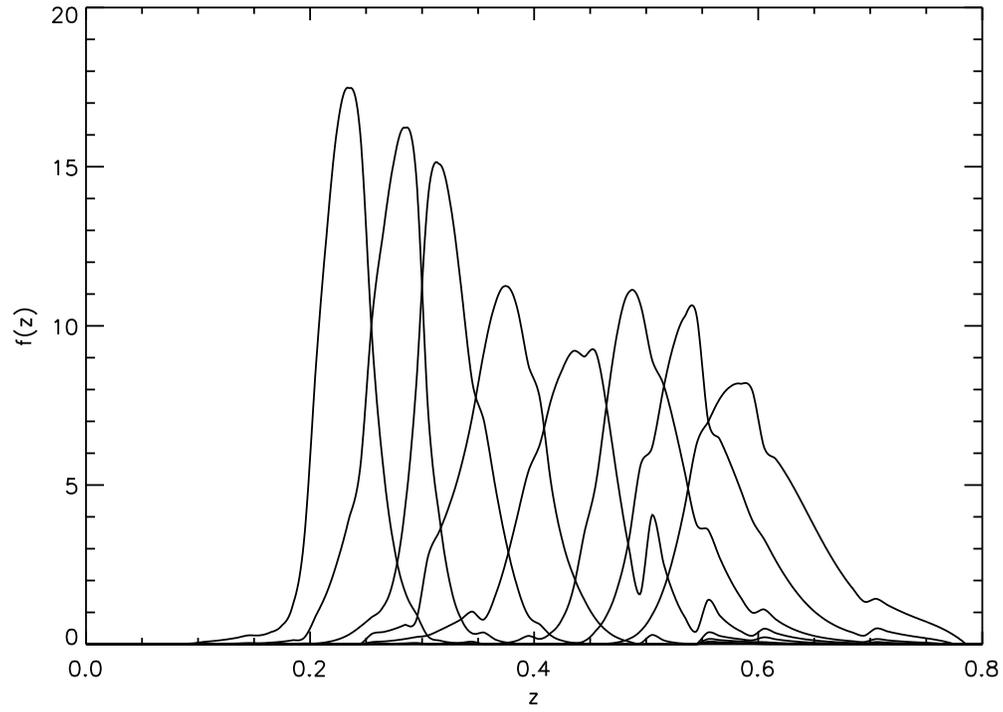}}}
\caption{The redshift distributions for the 8 photometric redshift slices.\label{F:redplot}}
\end{figure}

We use a sample of photometrically classified luminous red galaxies.  LRGs are the most luminous galaxies in the universe, making them appealing for probing maximal 
volume.  They also tend to be old stellar systems with uniform spectral energy distributions and a strong discontinuity at 4000 $\angs$, which enables precise 
photometric redshifts and hence measurements of $\{g_{2M}\}$ in multiple redshift slices.  This both reduces statistical error bars and allows tests for consistency.  
The cuts that define the photometric LRG sample are enumerated in Ref.~\cite{Padmanabhan:2004ic}.  The sample is divided into 8 photometric redshift slices of 
thickness $\Delta z_{\rm p}=0.05$ ranging from $z_{\rm p,min}=0.2$ to $z_{\rm p,max}=0.6$, using the ``single-template'' photo-$z$ algorithm of 
Ref.~\cite{Padmanabhan:2004ic}.  We plot the redshift distributions in Fig.~\ref{F:redplot}, while their properties are given in Table 
\ref{T:redshifts}.

\begin{figure}
\includegraphics[width=5.5in]{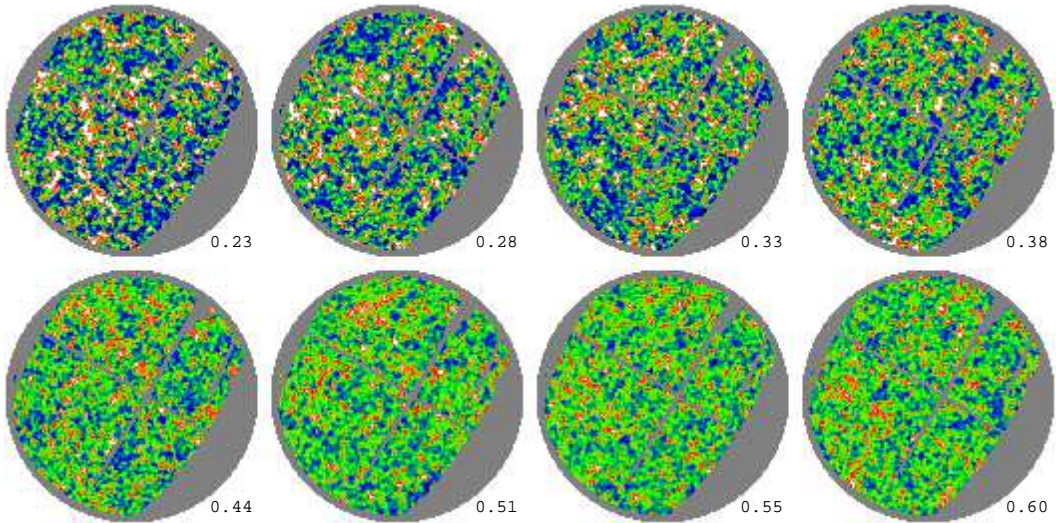}
\caption{\label{F:lssmaps}The LRG density in the 8 photometric redshift slices.  The 45$^\circ$ radius caps are displayed in a Lambert Azimuthal Equal-Area 
Projection, with the North Galactic Pole at the centre, $l=0^\circ$ at right, and $l=90^\circ$ at bottom.  The labels indicate the characteristic redshift of
each slice.}
\end{figure}

Our galaxy catalogue is a subset of that used by Ref.~\cite{Ho:2008bz}, but we restrict our attention to Galactic latitudes $b>45^\circ$.  This was done in 
Ref.~\cite{Padmanabhan:2006ku} to prevent stellar contamination in the data, and we decided to use the same cut accordingly.  We pixelize these galaxies as a number 
overdensity, $\delta_g=(n-\overline{n})/\overline{n}$, onto a HEALPix pixelization \cite{Gorski:2004by} of the sphere, with $1,418,213$ pixels.  This corresponds to a 
solid angle of $4662$ deg$^2$ (as opposed to 3528 used in Ref.~\cite{Padmanabhan:2006ku}).  The LRG maps thus generated are shown in Fig.~\ref{F:lssmaps}.

\begin{table}
\caption{\label{T:redshifts} Properties of the 8 redshift slices; $z_{\rm p}$ is the photometric redshift range, and $z_{\rm mean}$ is the mean (true) redshift of the 
slice.  $N_{gal}$ 
is the number of galaxies in the redshift slice, and $b_g$ is the linear bias.}
\begin{indented}
\item[]\begin{tabular}{@{}ccccc}
\br Label&$z_{\rm p}$&$z_{\rm mean}$&$N_{gal}$&$b_g$\\
                                  \mr z00&0.20-0.25&0.233&30812&1.74\\
                                  z01&0.25-0.30&0.276&38168&1.52\\
                                  z02&0.30-0.35&0.326&37963&1.67\\
                                  z03&0.35-0.40&0.376&55951&1.94\\
                                  z04&0.40-0.45&0.445&77798&1.75\\
                                  z05&0.45-0.50&0.506&138901&1.73\\
                                  z06&0.50-0.55&0.552&126318&1.80\\
                                  z07&0.55-0.60&0.602&93973&1.85\\ \br
\end{tabular}\end{indented}
\end{table}

\section{Formalism and estimators} \label{S:form}

\subsection{Galaxy density projections on the sky} \label{S:formproj}

We relate angular correlations in the sky to the direction-dependent matter power spectrum $P(\veck)$, which follows the derivation in Ref.~\cite{Padmanabhan:2006ku} 
for the statistically isotropic case.  A photometric galaxy survey measures the 2-dimensional projected galaxy overdensity $\delta_g(\hatn)$, which is related to the 
full 3-dimensional density via
\begin{eqnarray} \label{E:proj}
\delta_g(\hatn) = \int {\rm d}\chi f(\chi)\delta_g(\vecx=\chi\hatn)\, ,
\end{eqnarray}
where $\chi$ is the comoving distance, and $f(\chi)$ is the radial selection function, which is the normalized redshift distribution for the redshift slice.  (We 
leave out redshift space distortions here, but include them below.)
By Fourier transforming $\delta_g(\vecx)$ and using the identity
\begin{eqnarray} \label{E:expsum}
{\rm e}^{-{\rm i}\veck\cdot\hatn\chi} = \sum_{l=0}^\infty (2l+1){\rm i}^lj_l(k\chi)P_l(\hatk\cdot\hatn)\, ,
\end{eqnarray}
we obtain
\begin{eqnarray} \label{E:deltaexpand}
\delta_g(\hatn) = \sum_{l=0}^\infty {\rm i}^l(2l+1)\int\frac{{\rm d}^3k}{(2\pi)^3}P_l(\hatk\cdot\hatn)\delta_g(\veck)W_l(k)\, ,
\end{eqnarray}
where
\begin{eqnarray} \label{E:wl}
W_l(k) = \int {\rm d}\chi f(\chi)j_l(k\chi)
\end{eqnarray}
is the window function, and $j_l(x)$ and $P_l(x)$ are the $l^{\mbox{th}}$-order spherical bessel functions and Legendre polynomials, respectively.

The statistical properties of the 2-dimensional galaxy field can be derived in analogy to those for the CMB \cite{Pullen:2007tu}, with $W_l(k)$ replacing the CMB 
radiation transfer function $\Theta_l(k)$.\footnote{Note that with our definition of the Fourier transform, there is a relative factor of $(2\pi)^3$ between some of 
our formulas and those of Ref.~\cite{Pullen:2007tu}.}  If statistical isotropy is valid, 
then the two-point galaxy correlation function can be written as
\begin{eqnarray} \label{E:cordef}
C_g(\hatn,\hatn')|_{\rm SI} = \VEV{\delta_g(\hatn)\delta_g(\hatn')}|_{\rm SI}
= \sum_l \frac{2l+1}{4\pi} C_{g,l}P_l(\hatn\cdot\hatn')\, ,
\end{eqnarray}
where $C_{g,l}$ is the angular galaxy power spectrum:
\begin{eqnarray} \label{E:cl}
C_{g,l}=\frac{2}{\pi}\int_0^\infty {\rm d}k\,k^2\bar P_g(k)[W_l(k)]^2\, ,
\end{eqnarray}
where $\bar P_g(k)=b_g^2\bar P(k)$.  However, the presence of statistical anisotropy will cause additional terms to appear other than the 
ones in Eq.~\ref{E:cordef}.  Using Eqs.~(\ref{E:deltaexpand}) and (\ref{E:powerspectrum}), we find for the statistically anisotropic case
\begin{eqnarray} \label{E:covresult}
C_g(\hatn,\hatn') &=& \sum_l \frac{2l+1}{4\pi} C_{g,l}P_l(\hatn\cdot\hatn')
\nonumber \\ &&
+\sum_{LM}\sum_{lml'm'}D_{g,ll'}^{LM}X_{lml'm'}^{LM}R_{lm}(\hatn)R_{l'm'}(\hatn')\, .
\end{eqnarray}
Here, the set of $C_{g,l}$s are given by
the usual galaxy power spectrum for the case of statistical isotropy.  Statistical anisotropy produces the second term, where
\begin{eqnarray}\label{E:dll}
D_{g,ll'}^{LM}={\rm i}^{l-l'}\frac{2}{\pi}\int_0^\infty {\rm d}k\,k^2\bar P_g(k)g_{LM}W_l(k)W_{l'}(k)\, ,
\end{eqnarray}
and
\begin{eqnarray}\label{E:xlm}
X_{lml'm'}^{LM} = \int d^2\hatk\,R_{lm}(\hatk)R_{l'm'}(\hatk)R_{LM}(\hatk)\,
\end{eqnarray}
is the real 3-harmonic coupling coefficient.  Its expression in terms of Wigner 3-$j$ symbols is given in \ref{A:anicoef}.  Parity implies that $X_{lml'm'}^{LM}$ is 
nonzero only for $l+l'-L$ even.  Since earlier in the Introduction we showed $L$ is even, this requires $l+l'$ to be even and $D_{g,ll'}^{LM}$ to be real.  
Equations~(\ref{E:cl}) and (\ref{E:dll}) agree with similar results in Ref.~\cite{Pullen:2007tu}, and $X_{lml'm'}^{LM}$ in Eq.~(\ref{E:xlm}) is analogous to 
$\xi_{lml'm'}^{LM}$ 
in Ref.~\cite{Pullen:2007tu}.  Throughout this paper, we use upper-case indices $LM$ for statistical anisotropies in the matter power spectrum, and lower-case indices 
$lm$ for random anisotropies in the galaxy distribution.

\begin{figure}
\includegraphics[width=70 mm]{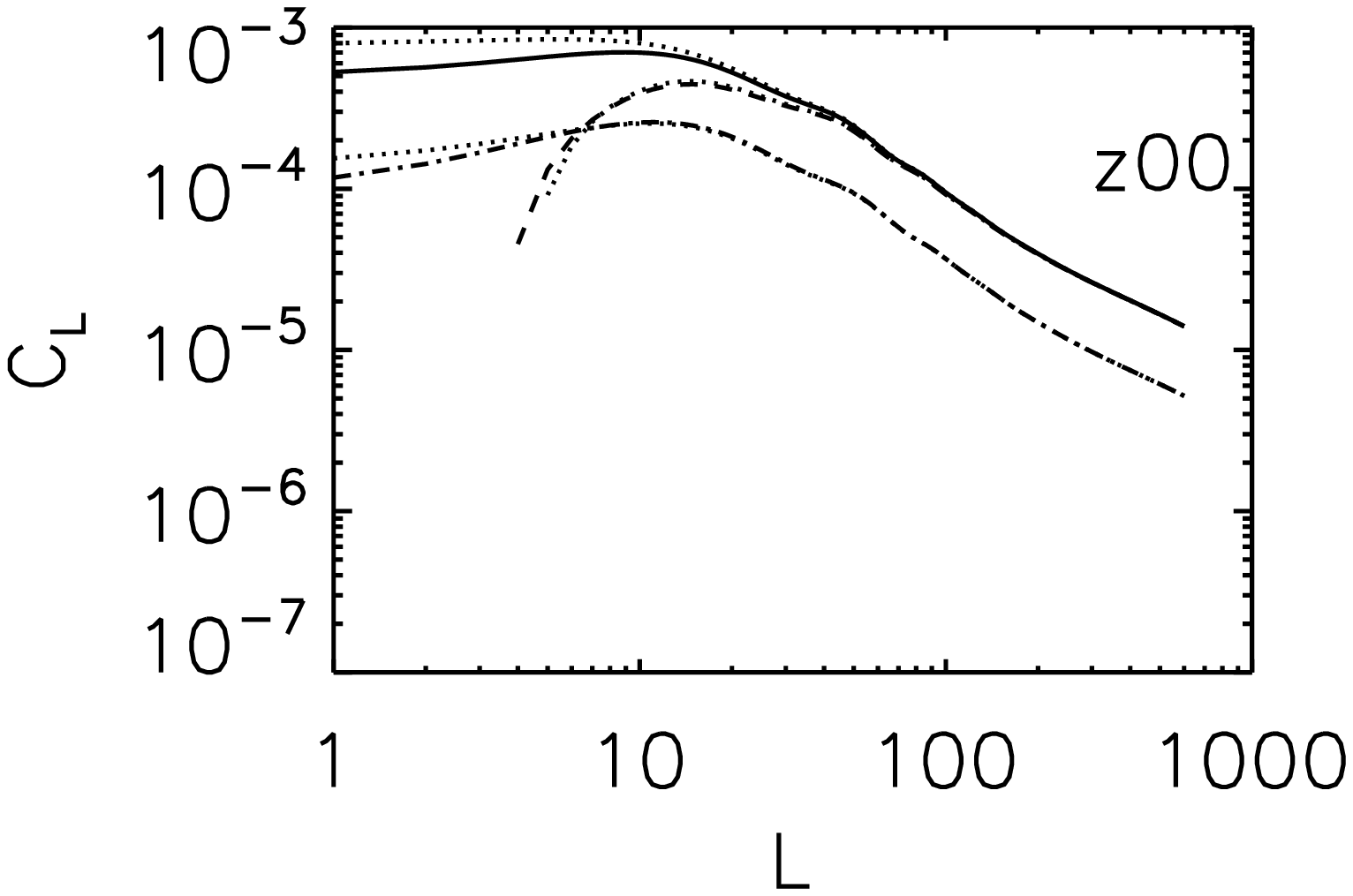}
\includegraphics[width=70 mm]{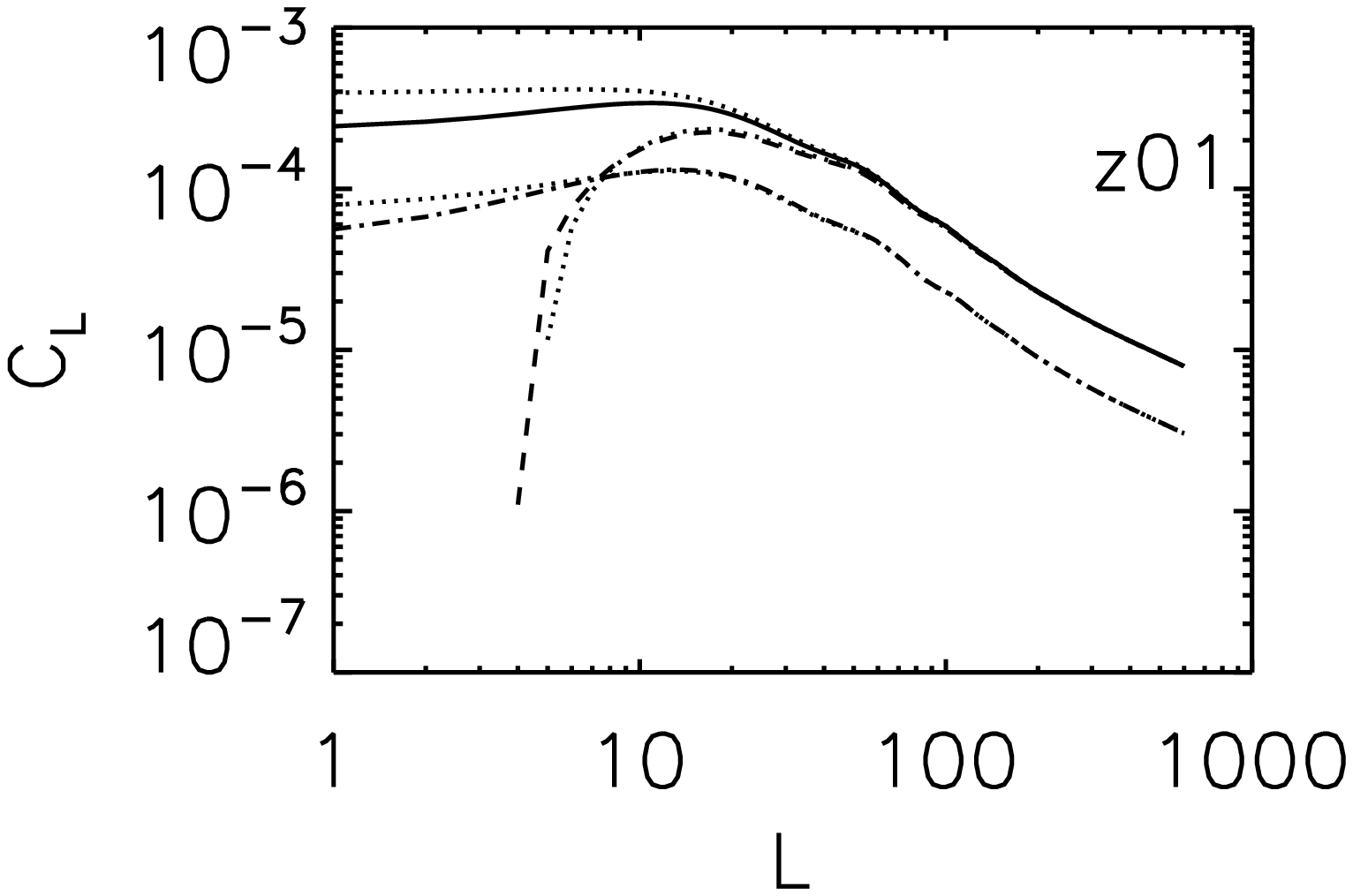}
\includegraphics[width=70 mm]{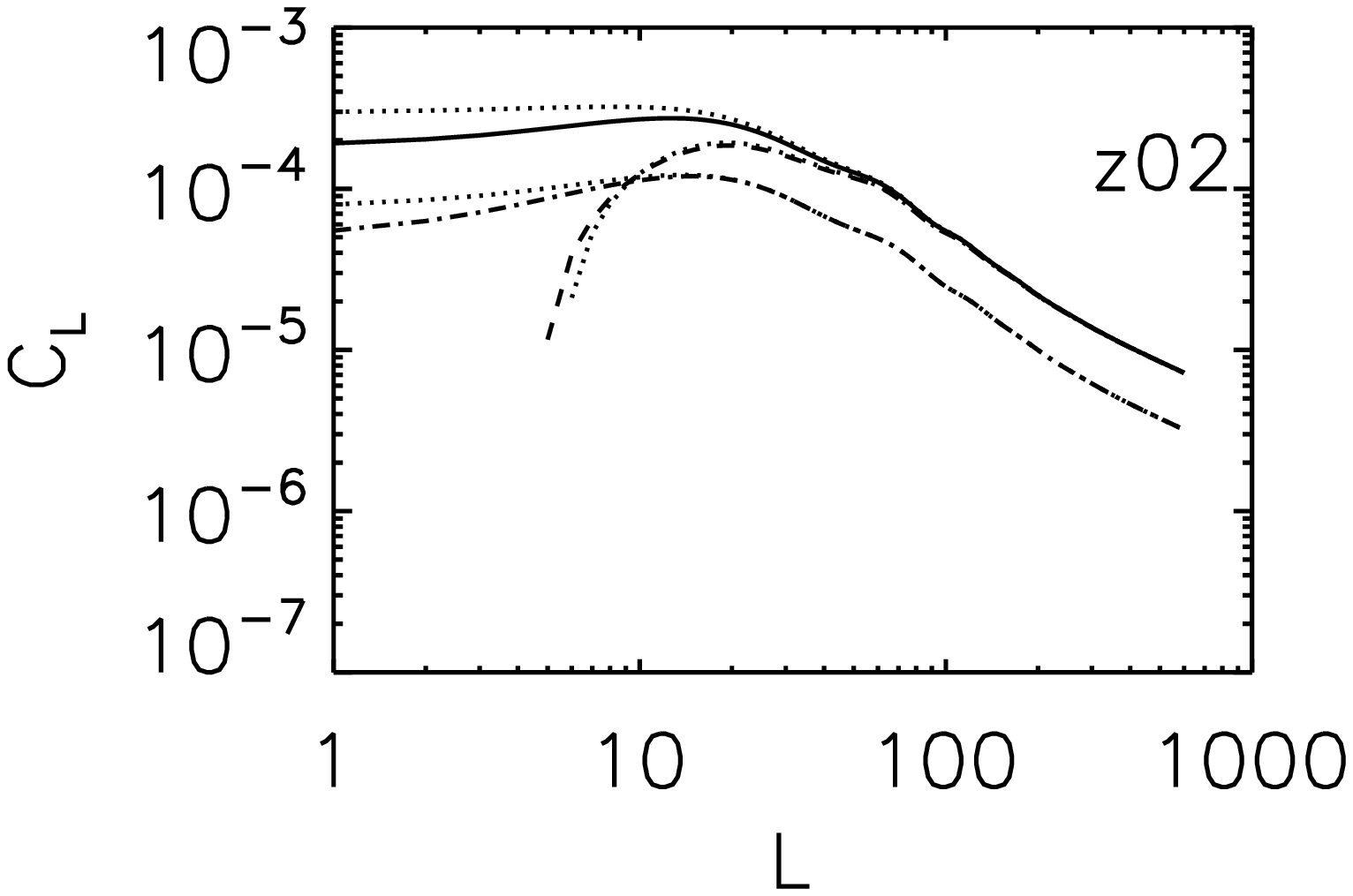}
\includegraphics[width=70 mm]{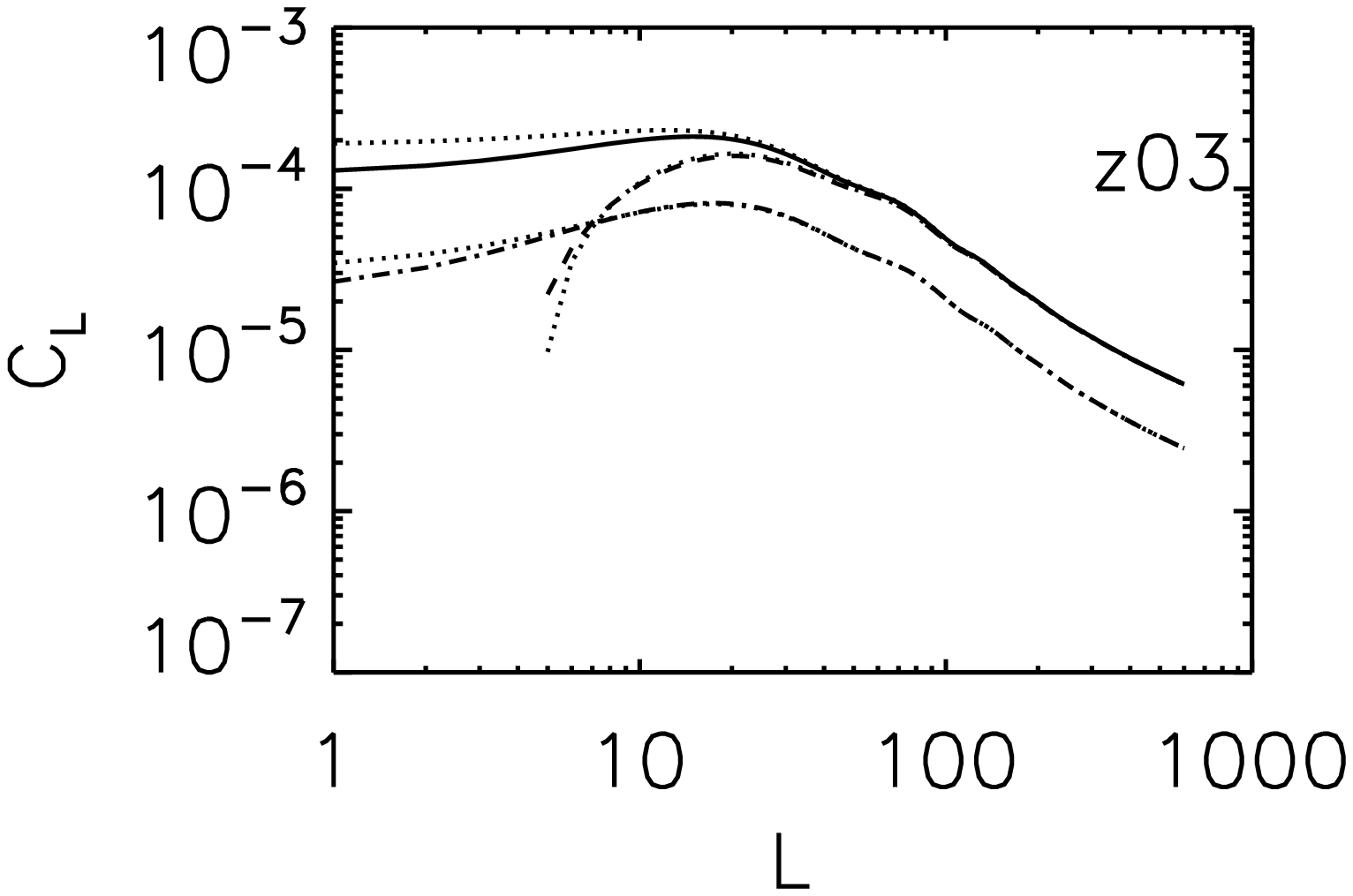}
\includegraphics[width=70 mm]{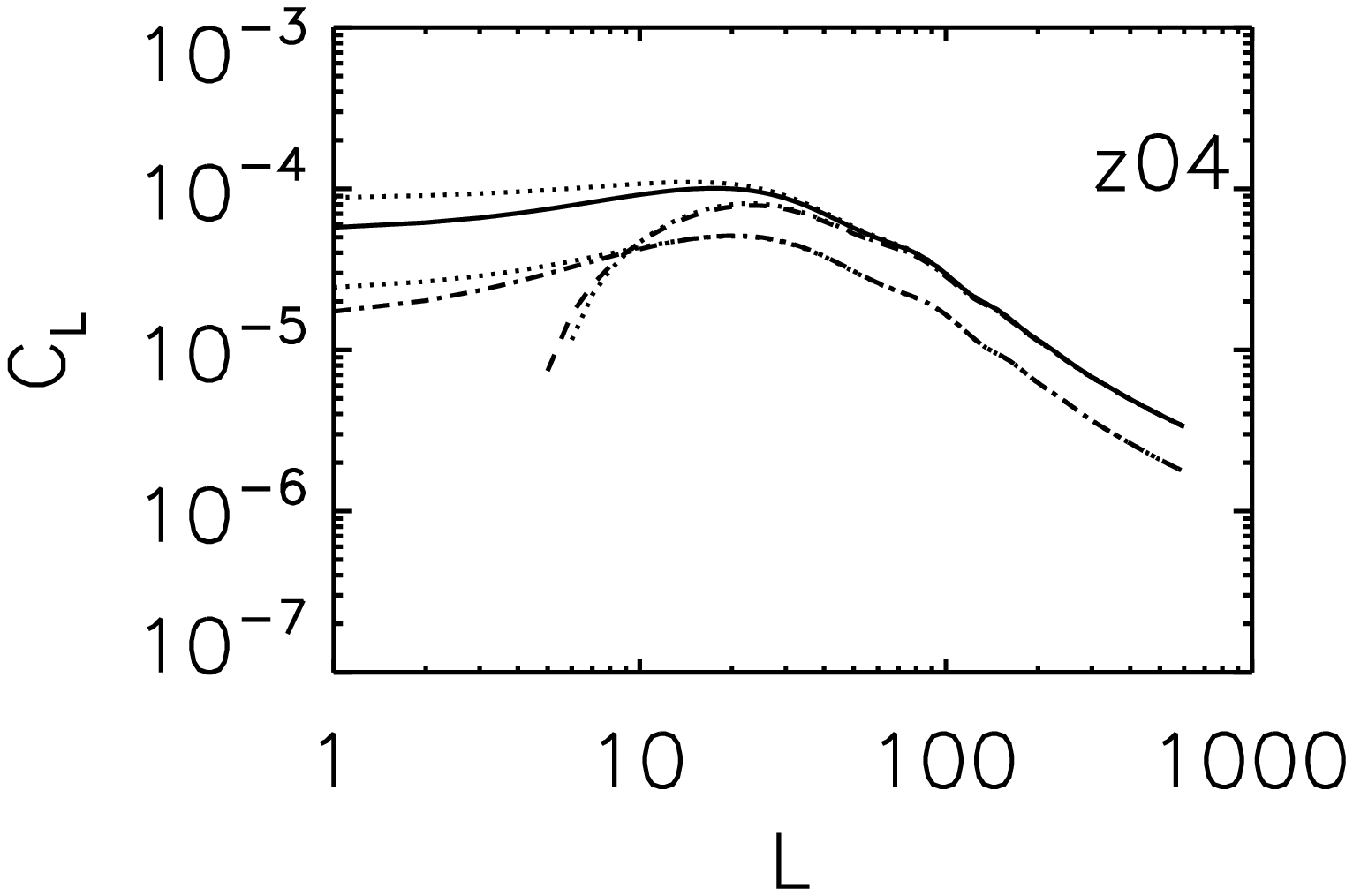}
\includegraphics[width=70 mm]{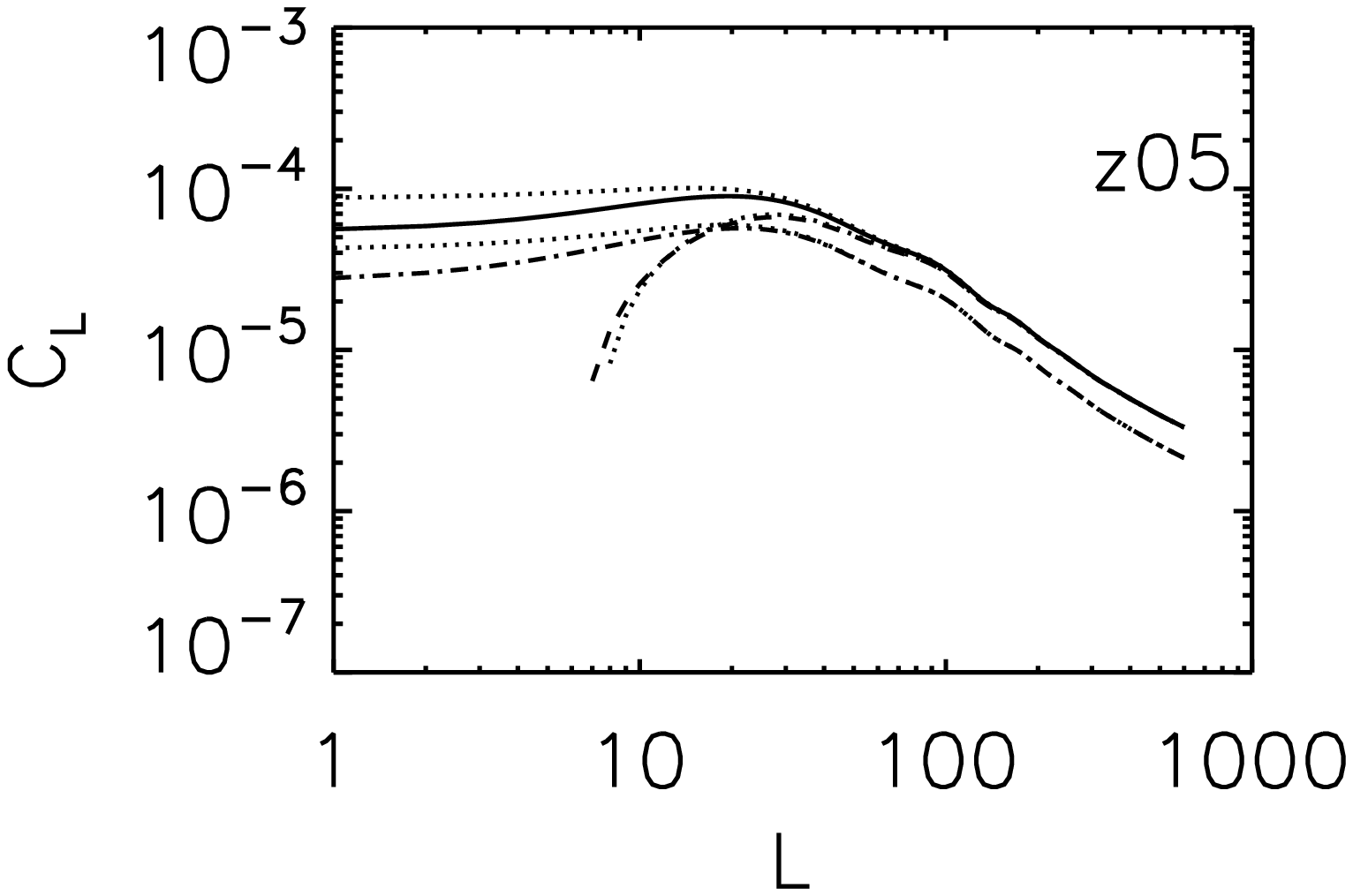}
\includegraphics[width=70 mm]{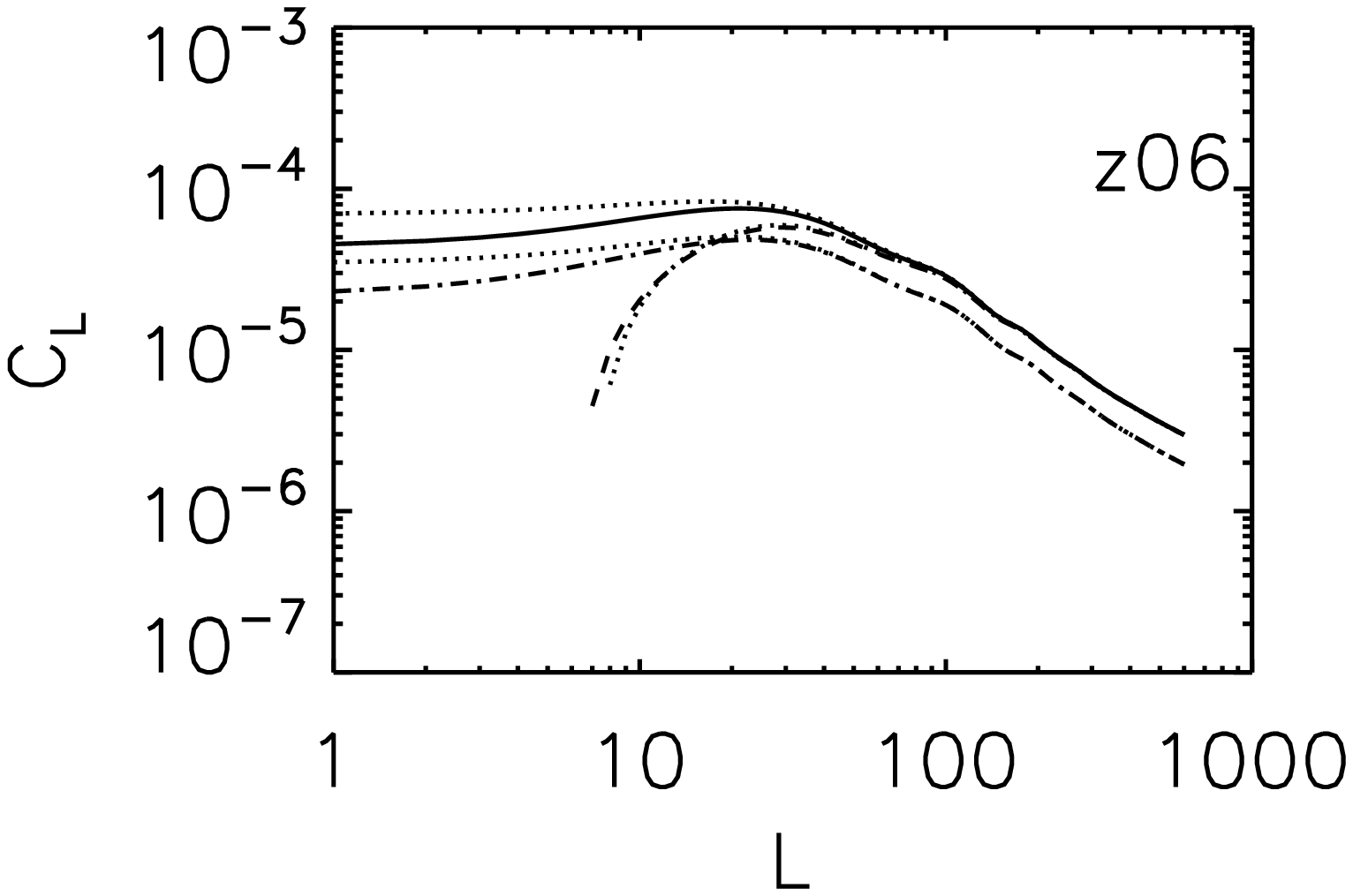}
\includegraphics[width=70 mm]{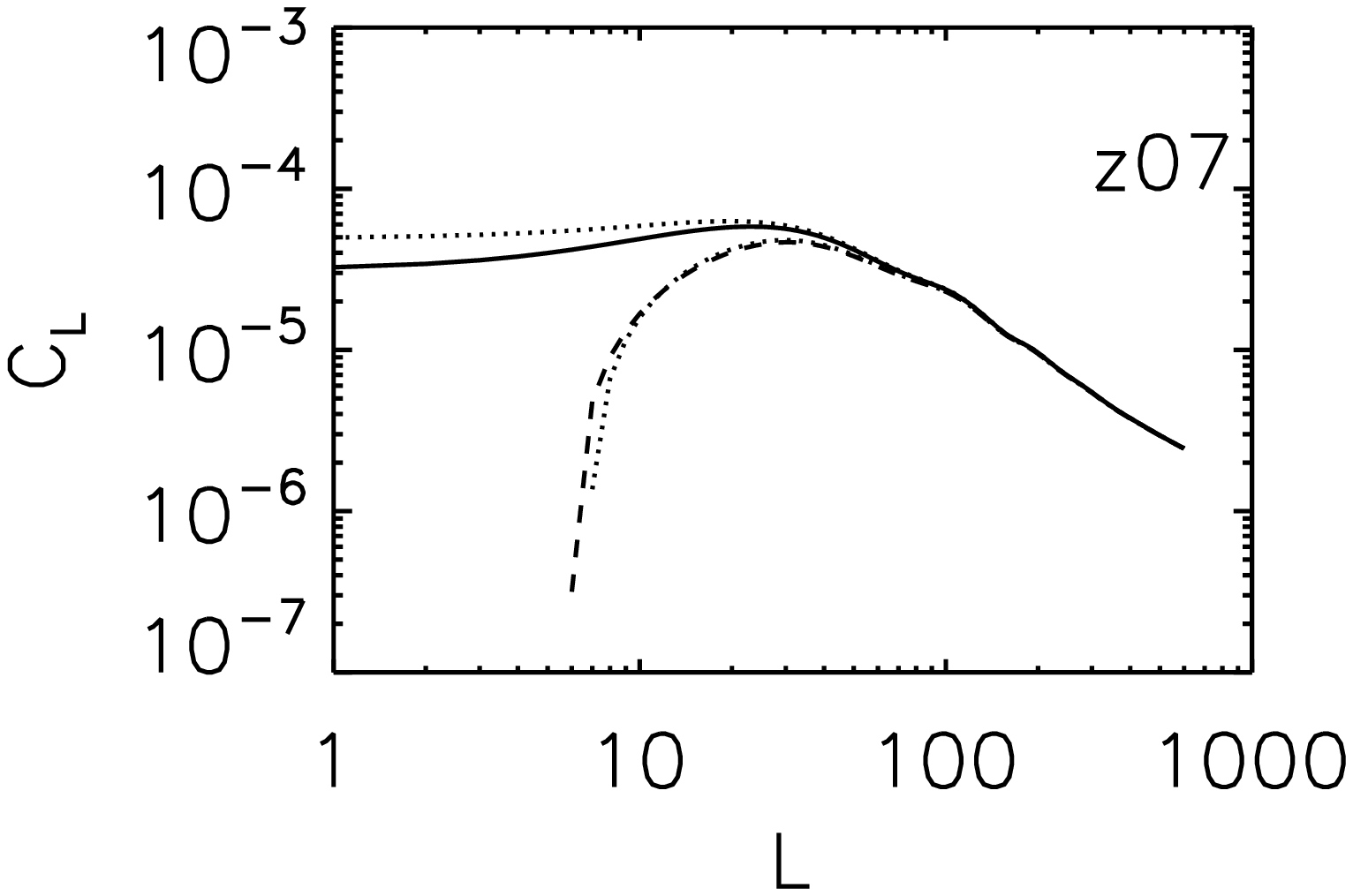}
\caption{The predicted angular power spectra for each of the 8 redshift slices.  The solid lines show nonlinear auto power spectra, while the dash-dotted lines show 
the cross power power spectra with the adjacent slice at higher redshift.  The dashed lines show the predicted $[-F_{g,l(l+2)}]$.  The dotted lines show the effect of 
redshift space distortions on all three spectra.\label{F:cls}}
\end{figure}

In Fig.~\ref{F:cls} we show the predicted angular galaxy power spectra $C_{g,l}$ for the eight redshift slices in our analysis assuming our fiducial cosmology.  We 
use {\sc CMBFast} \cite{cmbfast} to calculate $\bar P(k)$, and we use the {\sc Halofit} prescription \cite{Smith:2002dz} to evolve $\bar P(k)$ into the nonlinear 
regime.  Note that we do not attempt to account for the nonlinear evolution of $g_{LM}$, which only suppresses the primordial anisotropy by $\lesssim 7\%$ on 
quasilinear scales \cite{Ando:2008zza}.  We 
also display $[-F_{g,l(l+2)}]$, where $F_{g,ll'} = D_{g,ll'}^{LM}/g_{LM}$.  We only need to show $F_{g,l(l+2)}$ because we are interested only in quadrupolar 
statistical anisotropy, or $L=2$.  For this case it can be shown that $X_{lml'm'}^{2M}$ is zero except for the cases $l'=l$ and $l'=l\pm2$, and 
$F_{g,ll}=C_{g,l}$.\footnote{For nonzero $X_{lml'm'}^{2M}$, the triangle inequality requires $|l-l'|\le L=2$, and parity requires $l-l'$ to be even.}  
Notice that for large $l$ and smooth $f(\chi)$, $W_l(k)\rightarrow \sqrt{\pi/l}\,f(l/k)/(2k)$; then we have $W_{l+2}\simeq W_l$ and $F_{g,l(l+2)}\simeq -C_{g,l}$.

We also show in Fig.~\ref{F:cls} $C_{g,l}$ and $[-F_{g,l(l+2)}]$ when the effect of redshift space distortions is included.  We include this effect by substituting for the window function $W_l = W_l^0+W_l^r$, where $W_l^0$ is the window function shown in Eq.~\ref{E:wl} and $W_l^r$ is given by
\begin{eqnarray} \label{E:wlr}
W_l^r(k) &=&\beta\Biggl[\frac{2l^2+2l-1}{(2l+3)(2l-1)}W_l^0(k)\nonumber\\&&
\hphantom{\beta\biggl[}-\frac{l(l-1)}{(2l-1)(2l+1)}W_{l-2}^0(k)\nonumber\\&&
\hphantom{\beta\biggl[}-\frac{(l+1)(l+2)}{(2l+1)(2l+3)}W_{l+2}^0(k)\Biggr],
\end{eqnarray}
and $\beta$ is the redshift distortion parameter given approximately by $\beta \sim \Omega_m^{0.6}/b_g$.  The formulas for redshift space distortions in the angular 
galaxy power spectrum were derived in Ref.~\cite{Padmanabhan:2006ku}.

\subsection{Estimation of power and statistical anisotropy} \label{S:formest}

We construct a quadratic estimator \cite{Tegmark:1996qt, Padmanabhan:2002yv} to measure the anisotropy coefficients $g_{2M}$.  As always when searching 
for anisotropies, it is necessary to fit simultaneously for the galaxy power spectrum, the anisotropy, and any systematics terms that may be present in the data.  The 
basic premise is to write the galaxy density fluctuation map as a vector $\mathbf{\delta_g}$ of length $N_{\rm pix}$.  This vector has an $N_{\rm pix}\times N_{\rm 
pix}$ covariance matrix $\mathbf{C}$, which we parameterize as
\begin{eqnarray}
\mathbf{C} = \sum_{i=1}^{N_t} p_i \mathbf{C}_{,i},
\end{eqnarray}
where the $\{p_i\}_{i=1}^{N_t}$ are parameters to be estimated and $\{\mathbf{C}_{,i}\}_{i=1}^{N_t}$ are ``templates.''  The notation also serves to remind us that 
$\mathbf{C}_{,i} = \partial\mathbf{C}/\partial p_i$.  In the case of a Gaussian random field, the minimum-variance unbiased quadratic estimators for the $\{p_i\}$ are
\begin{eqnarray} \label{E:param}
\mathbf{\hat{p}}=\mathbf{F}^{-1}\mathbf{q}\, ,
\end{eqnarray}
where
\begin{eqnarray} \label{E:fisher}
F_{ij}=\frac{1}{2}\rm{tr}\left[\mathbf{C},_i\mathbf{w}\mathbf{C},_j\mathbf{w}\right]
\end{eqnarray}
is the Fisher matrix and
\begin{eqnarray} \label{E:quad}
q_i = \frac{1}{2}\mathbf{\delta_g^T}\mathbf{w}\mathbf{C},_i\mathbf{w}\mathbf{\delta_g}\,.
\end{eqnarray}
Here $\mathbf{w}$ is a weighting matrix, which should be taken equal to the inverse of the covariance matrix $\mathbf{C}$.  Fortunately, the estimator 
$\mathbf{\hat{p}}$ remains unbiased (but not necessarily minimum-variance) for any choice of weight $\mathbf{w}$, and regardless of whether the true galaxy field 
$\delta_g$ is Gaussian or not.  For our analysis, we take the weight to be $\mathbf{w}=(\mathbf{S}+\mathbf{N})^{-1}$, where $\mathbf{S}$ is the signal covariance 
matrix (diagonal in $lm$-space, and using the theoretical power spectra) and $\mathbf{N}$ is the Poisson noise.  The matrix inversion and trace estimation are done by 
the iterative and stochastic methods described in detail in Refs.~\cite{Hirata:2004rp, Padmanabhan:2006ku}.

We next turn our attention to the template construction.  The simplest template is the Poisson noise itself,
\begin{eqnarray} \label{E:noise}
C_{ij,N} = \frac{\delta_{ij}}{\overline{n}}\,,
\end{eqnarray}
where $\overline{n}$ is the mean number of galaxies per pixel, and $N$ is the noise amplitude (1 for pure Poisson noise).
We may also parameterize the isotropic part of the power spectrum by band power amplitudes $\tilde{C}_n$ with $C_{g,l}=\sum_{n=1}^{N_{\rm bin}}\tilde{C}_n\eta_l^n$,
where $\eta_l^n$ is a step function that is 1 when $l$ is in bin $n$ and 0 otherwise.  The corresponding template is
\begin{eqnarray} \label{E:clsum}
\frac{\partial C_{ij}}{\partial \tilde{C}_n} = \sum_{lm} R_{lm}(\hatn_i) R_{lm}(\hatn_j) \eta_l^n.
\end{eqnarray}
We use 18 bins in $l$, ranging from $l=2$ up through 600.

For the anisotropy parameters $g_{2M}$, the templates are somewhat more complicated.  We first extract $g_{2M}$ from the anisotropy amplitude $D_{g,ll'}^{2M}$:
\begin{eqnarray}\label{E:dgf}
D_{g,ll'}^{2M}=g_{2M}F_{g,ll'}\, ,
\end{eqnarray}
where
\begin{eqnarray}\label{E:Fll}
F_{g,ll'}={\rm i}^{l-l'}\frac{2}{\pi}\int_0^\infty {\rm d}k\,k^2\bar P_g(k)W_l(k)W_{l'}(k)\, .
\end{eqnarray}
Inspection of Eq.~\ref{E:covresult} then leads us to:
\begin{eqnarray} \label{E:covderiv}
\frac{\partial C_{ij}}{\partial g_{2M}}&=&\sum_{lmm'}\Biggl[C_{g,l}X_{lmlm'}^{2M}R_{lm}(\hatn_i)R_{lm'}(\hatn_j)
\nonumber \\ &&
+F_{g,l(l+2)}X_{lm(l+2)m'}^{2M}R_{lm}(\hatn_i)R_{(l+2)m'}(\hatn_j)
\nonumber\\
&&+F_{g,l(l-2)}X_{lm(l-2)m'}^{2M}R_{lm}(\hatn_i)R_{(l-2)m'}(\hatn_j)\Biggr]\,.
\end{eqnarray}

In addition to these ``essential'' templates, we have also added two others to project out various systematics that could mimick statistical anisotropy.  In 
particular, if there are photometric calibration errors that vary slowly across the survey (either coloured or grey)\footnote{Coloured errors apply to an error in 
the relative calibration of different bands, e.g. $g-r$, whereas grey errors leave colours fixed but vary the magnitude of an object.}, then the depth or effective 
redshift 
may vary, which would lead to a modulation of both the signal power spectrum and the noise level across the sky.

We model the modulation of the signal power spectrum by considering a modulation in the fractional density perturbation field in the form 
$\delta'(\hatn)=[1+h(\hatn)]\delta(\hatn)$.  This modulation will cause the 
two-point galaxy correlation function to have an extra factor of $[1+h(\hatn)][1+h(\hatn')]\simeq1+h(\hatn)+h(\hatn')$, assuming $h(\hatn)\ll1$.  We choose to allow 
$h(\hatn)$ to have a quadrupole pattern.  (Other forms of slow modulation across the sky, e.g. a dipole, should be degenerate with a quadrupole given that our data is 
only in a cap of radius 45$^\circ$.)

In this case $\delta'(\hatn)$ can be written as
\begin{eqnarray} \label{E:dh}
\delta'(\hatn)=\left[1+\sum_{M=-2}^2h_{2M}R_{2M}(\hatn)\right]\delta(\hatn)\, ,
\end{eqnarray}
where $h_{2M}$ are the modulation parameters.  In the case of modulation with no statistical anisotropy, $C_g(\hatn,\hatn')$ is given by
\begin{eqnarray} \label{E:covh}
C_g(\hatn,\hatn')= \left\{1+\sum_{M=-2}^2h_{2M}\left[R_{2M}(\hatn)+R_{2M}(\hatn')\right]\right\}C_g(\hatn,\hatn')|_{SI}\, ,
\end{eqnarray}
where $C_g(\hatn,\hatn')|_{SI}$ is given by Eq.~(\ref{E:cordef}).  By using the identity
\begin{eqnarray}
R_{lm}(\hatn)R_{l'm'}(\hatn)=\sum_{LM}X_{lml'm'}^{LM}R_{LM}(\hatn)\, ,
\end{eqnarray}
we find
\begin{eqnarray} \label{E:covhsi}
C_g(\hatn,\hatn')&=& C_g(\hatn,\hatn')|_{\rm SI}
\nonumber \\ &&
+\!\!\sum_{lml'm'M}\! h_{2M}\left(C_{g,l}+C_{g,l'}\right)X_{lml'm'}^{2M}R_{lm}(\hatn)R_{l'm'}(\hatn'),
\end{eqnarray}
hence
\begin{eqnarray} \label{E:h}
\frac{\partial C_{ij}}{\partial
h_{2M}}&=&\sum_{lmm'}\Biggl[2C_{g,l}X_{lmlm'}^{2M}R_{lm}(\hatn_i)R_{lm'}(\hatn_j)
\nonumber \\ &&
+(C_{g,l}+C_{g,l+2})X_{lm(l+2)m'}^{2M}R_{lm}(\hatn_i)R_{(l+2)m'}(\hatn_j)
\nonumber\\
&&+(C_{g,l}+C_{g,l-2})X_{lm(l-2)m'}^{2M}R_{lm}(\hatn_i)R_{(l-2)m'}(\hatn_j)\Biggr].
\end{eqnarray}
An analogous construction for modulation of the Poisson noise gives
\begin{eqnarray} \label{E:noisecovderiv}
\frac{\partial C_{ij}}{\partial f_{2M}} = \frac{\delta_{ij}}{\overline{n}}R_{2M}(\hatn_i)\, .
\end{eqnarray}
These 10 parameters ($h_{2M}$ and $f_{2M}$) are jointly estimated with $\{\tilde C_n,g_{2M}\}$.

\subsection{Gaussian simulations} \label{S:test}

We test our estimator on a suite of simple simulations in order to verify its ability to detect anisotropy when it is present (and to measure zero when anisotropy 
is not present).  Gaussian simulations are sufficient for this purpose since a quadratic estimator by construction cannot be sensitive to higher moments of the data.

We perform two tests, one 
without anisotropy or modulation and one with both.  In each test, we use the power spectrum $C_{g,l}$ and $F_{g,l(l+2)}$ of redshift slice z00 for our fiducial 
cosmology to construct 50 sets of simulated galaxy perturbation maps over the pixels in our analysis' viewing region.  We also add Gaussian noise to each pixel, with 
the variance in the noise set equal to $1/\overline{n}_{\rm gal}$, where $\overline{n}_{\rm gal}$ is the average number of galaxies per pixel for the redshift slice 
(consistent with Poisson fluctuations).  Then, we run each simulation through the algorithm to find an estimated set of parameters $\tilde{C}_n$, $g_{2M}$, and 
$h_{2M}$.  We then average these parameters over all 50 simulations to find an output set of parameters, which we then compare to our input parameters for 
constructing the simulations.  Since we do not input $\tilde{C}_n$ directly, we instead compare the output $\tilde{C}_n$ to $C_{g,l}$ at the median $l$ of bin $n$.  
Note that the variance used to compare the input and output parameter sets is equal to the variance of one simulation, taken from the diagonal of the inverse fisher matrix, divided by the number of simulations.

\begin{figure}
{\scalebox{0.4}{\includegraphics{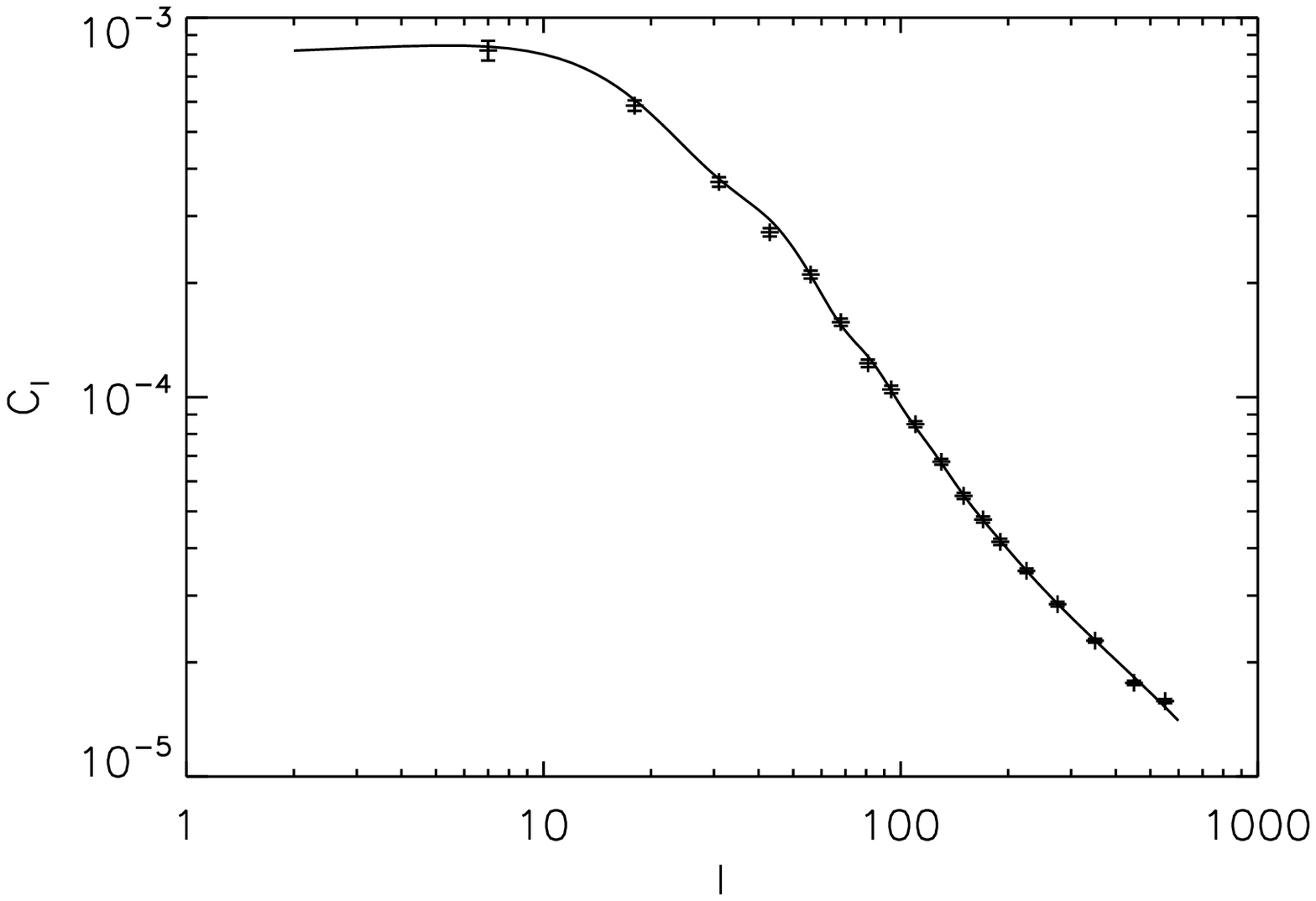}}
\scalebox{0.4}{\includegraphics{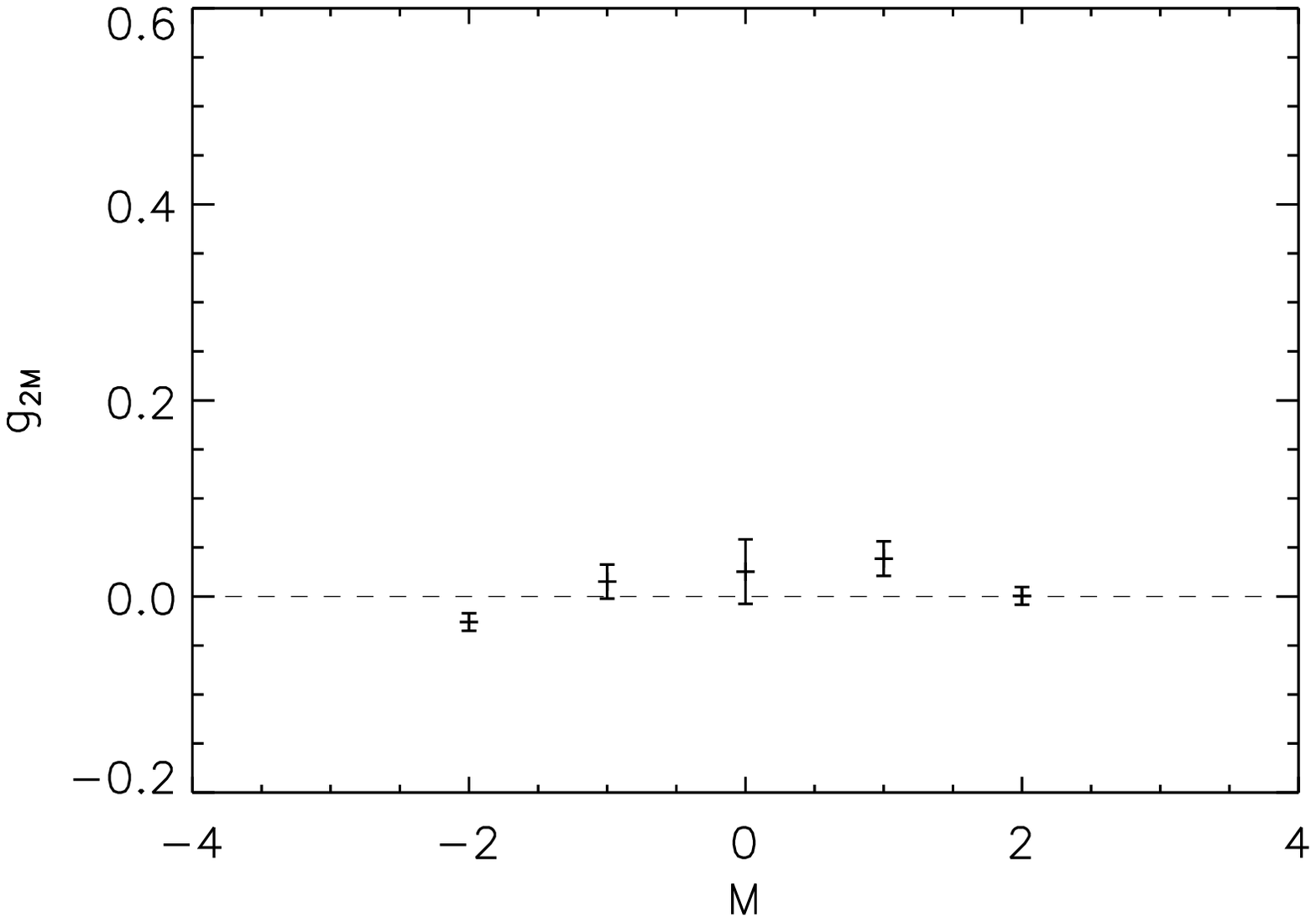}}
\scalebox{0.4}{\includegraphics{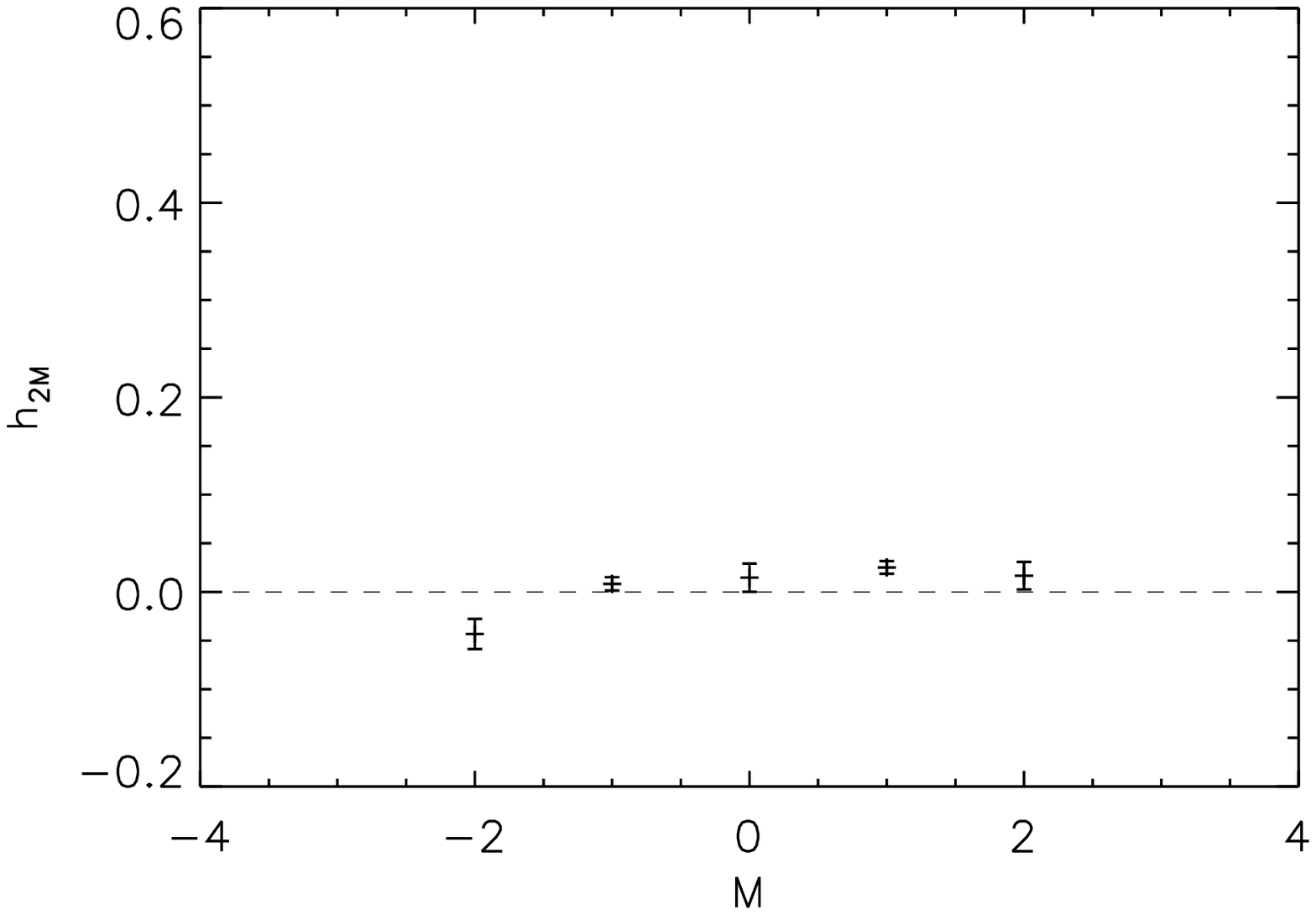}}
}
\caption{The parameter values for the simulation test with no input anisotropy or modulation, including 1-sigma errors.  The top left panel shows the $\tilde{C}_n$, the top right panel shows 
the 
$g_{2M}$s, and the bottom panel shows the $h_{2M}$s.\label{F:cgh0}}
\end{figure}

For the first test, our simulations had input parameters $g_{2M}=h_{2M}=0$ for all $M$.  Since there is no covariance between the simulated $\delta_{g,lm}$, the real 
spherical harmonic coefficients of $\delta_g(\hatn)$, we can simulate each $\delta_{g,lm}$ independently.  A plot of the input and output parameter values for 
$\tilde{C}_n$, $g_{2M}$, and $h_{2M}$ are shown in Fig.~\ref{F:cgh0}.  In the figures we see good agreement between the input and output values.  This test shows us 
that our algorithm should not see anisotropy or modulation where there is none.  We also see that the error for $g_{2,0}$ is larger than the errors for the other $g_{2M}$s.  This is due to a lack of data in the equatorial plane while $g_{2,0}$ parametrizes a quadrupole along the z-direction.  Our data set makes this type of quadrupole less distinguishable from a uniform excess over the whole sky than with other quadrupole types.

\begin{figure}
{\scalebox{0.4}{\includegraphics{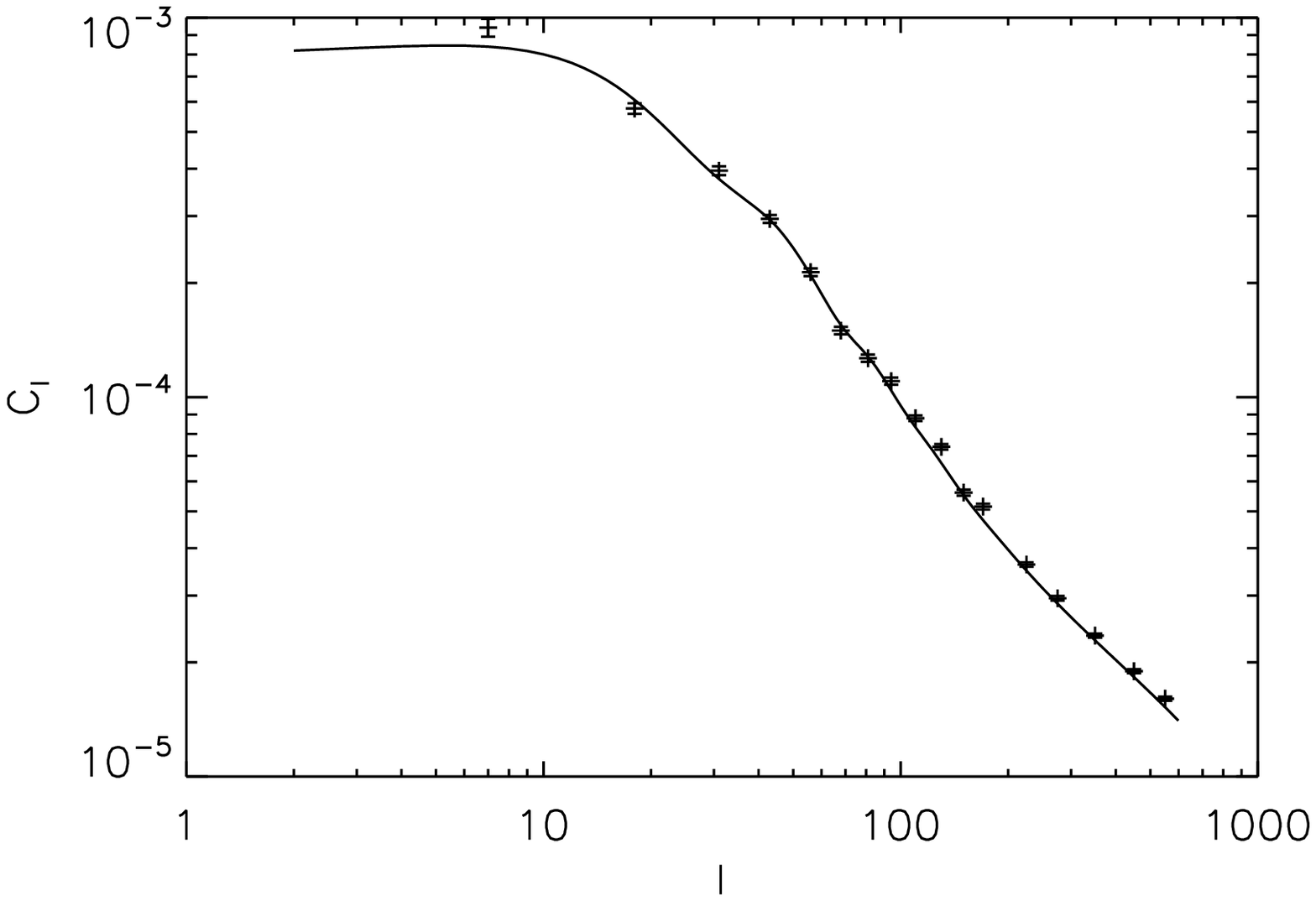}}
\scalebox{0.4}{\includegraphics{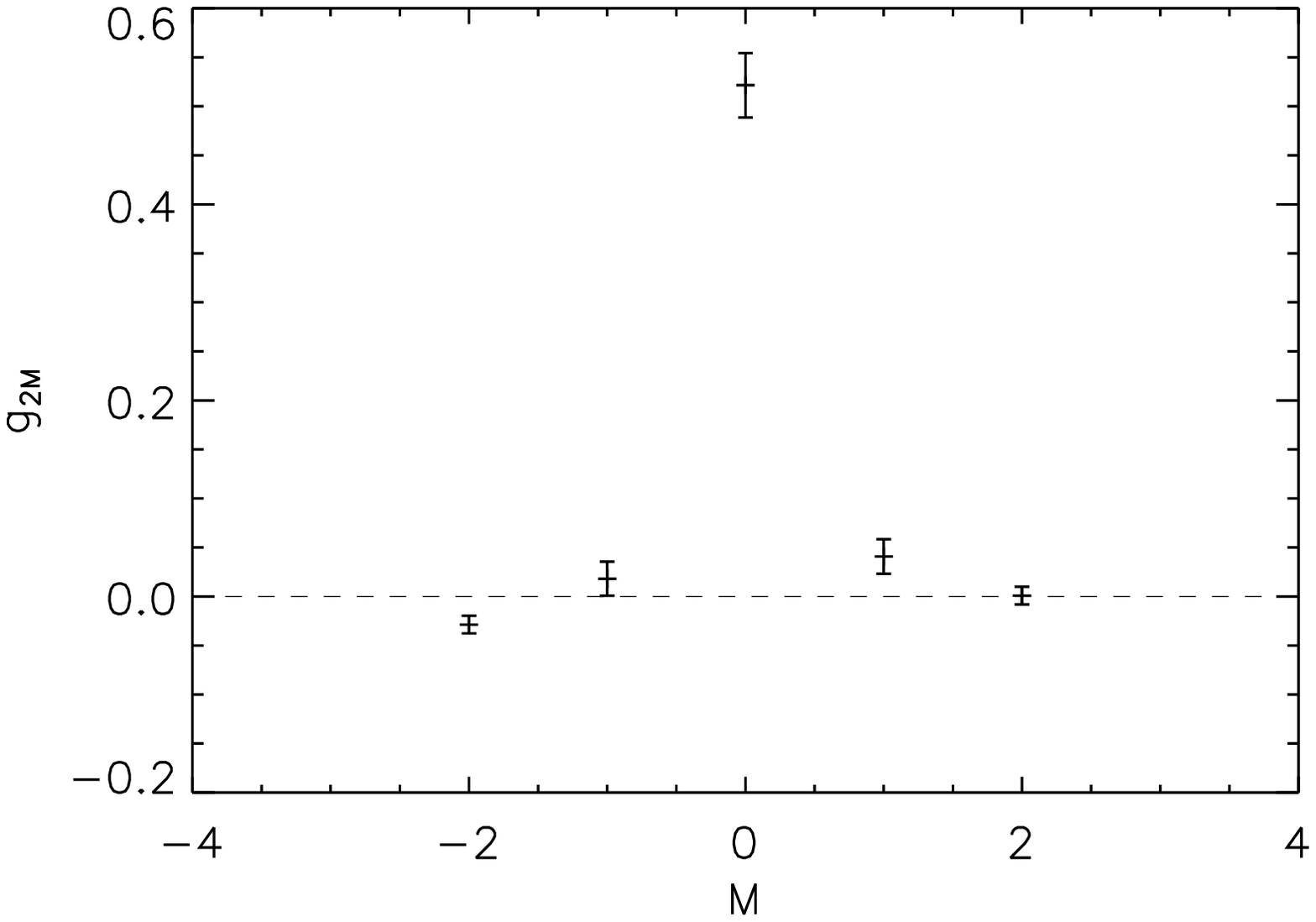}}
\scalebox{0.4}{\includegraphics{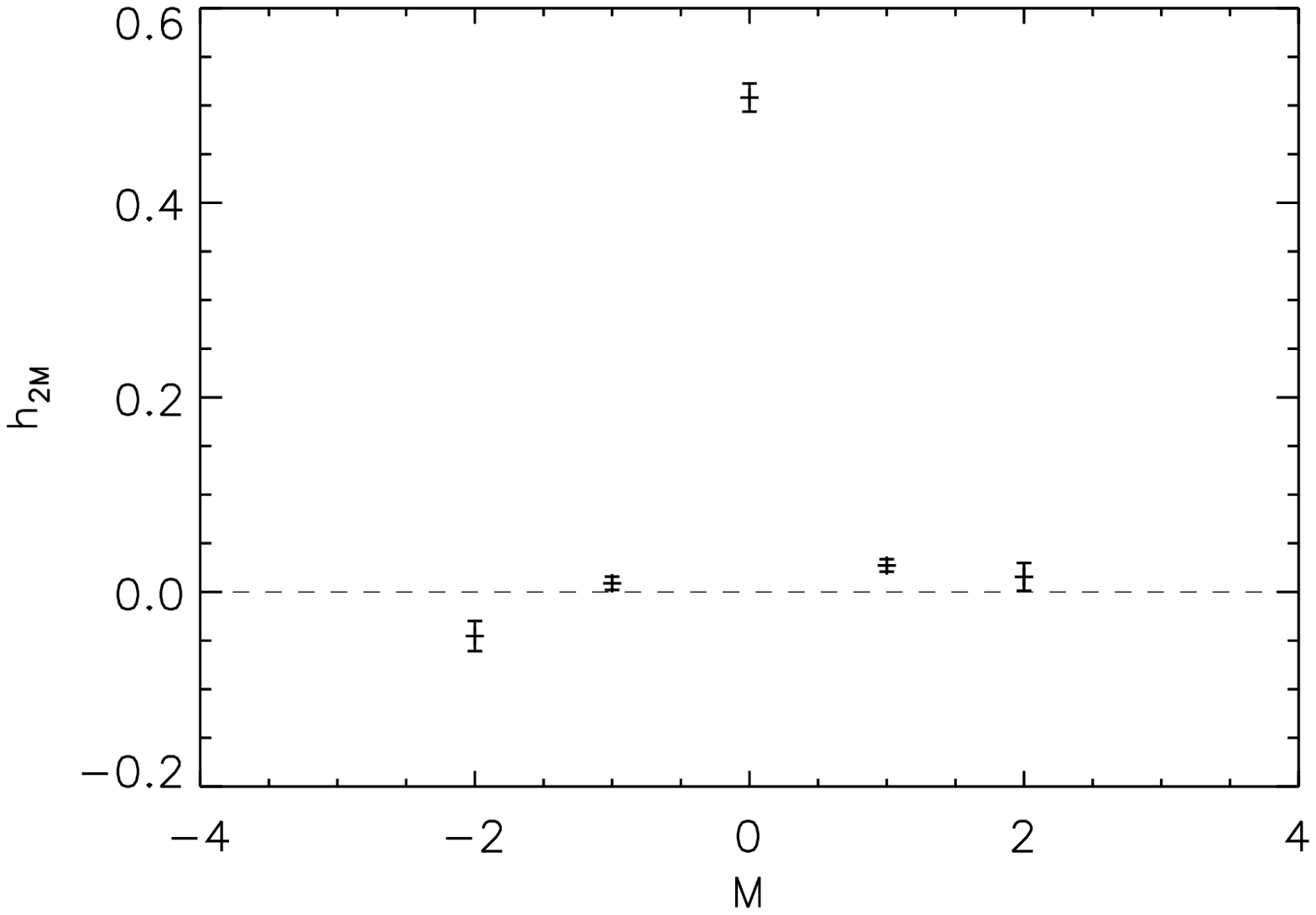}}
}
\caption{The parameter values for the simulation test with input anisotropy and modulation, including 1-sigma errors.  The top left panel shows the $\tilde{C}_n$, the top right panel shows 
the $g_{2M}$s, and the bottom panel shows the $h_{2M}$s.  Note that the input values for $g_{20}$ and $h_{20}$ are 0.5.\label{F:cgh}}
\end{figure}

In the second test, we simulate anisotropic power and modulation in the $z$-direction by setting $g_{20}=h_{20}=0.5$.  Now that the $\delta_{g,lm}$s are correlated, 
their simulation is no longer trivial.  To construct the simulation, we define the matrix $E_{ll'}^{(m)}$, equal to $\VEV{\delta_{g,lm}\delta_{g,l'm}}$ in the case 
$g_{2M}=h_{2M}=0$ for all $M$ except $M=0$.  This matrix is given by
\begin{eqnarray} \label{E:ell}
E_{ll'}^{(m)}=C_{g,l}\delta_{ll'}+g_{20}F_{g,ll'}X_{lml'm}^{20}+h_{20}(C_{g,l}+C_{g,l'})X_{lml'm}^{20}\, .
\end{eqnarray}
To construct our simulation, we perform a Cholesky decomposition on $\mathbf{E^{(m)}}$ to find the triangular matrix $\mathbf{L^{(m)}}$ such that 
$\mathbf{E}^{(m)}=\mathbf{L}^{(m)}\mathbf{L}^{(m)\,T}$.  We use $\mathbf{L^{(m)}}$ to construct $\mathbf{\delta_g^{(m)}}=\mathbf{L^{(m)}}\mathbf{x^{(m)}}$, where 
$\mathbf{x^{(m)}}$ is a Gaussian random matrix with zero mean and unit variance.  This setup gives us the desired covariances we need in the $\delta_{g,lm}$s.  A plot 
of the input and output parameter values for $\tilde{C}_n$, $g_{2M}$, and $h_{2M}$ in this case are shown in Fig.~\ref{F:cgh}.  In the figures we see good agreement 
between the input and output values, including for $g_{20}$ and $h_{20}$.

\section{Results} \label{S:result}

\begin{figure}
\includegraphics[width=70 mm]{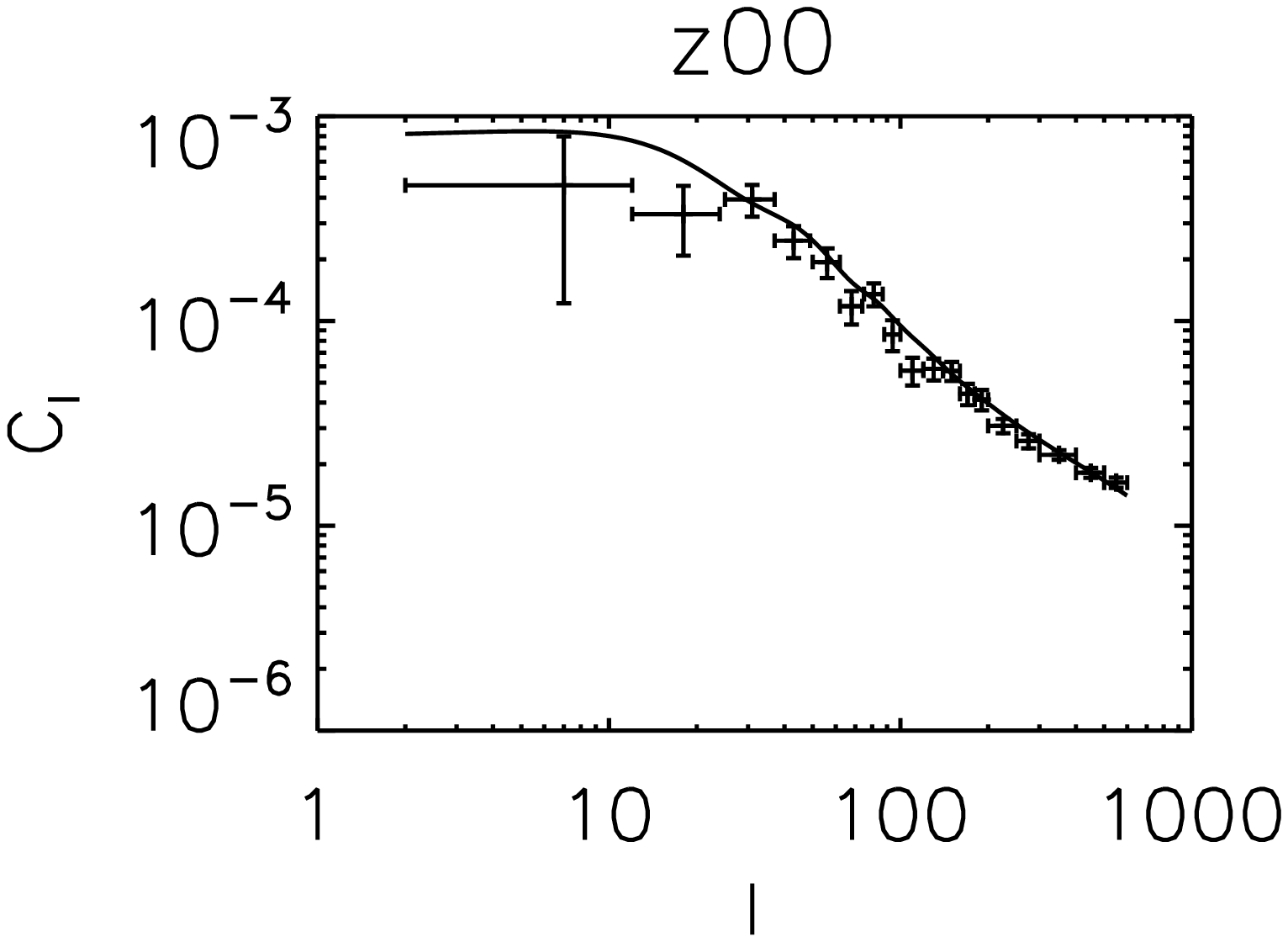}
\includegraphics[width=70 mm]{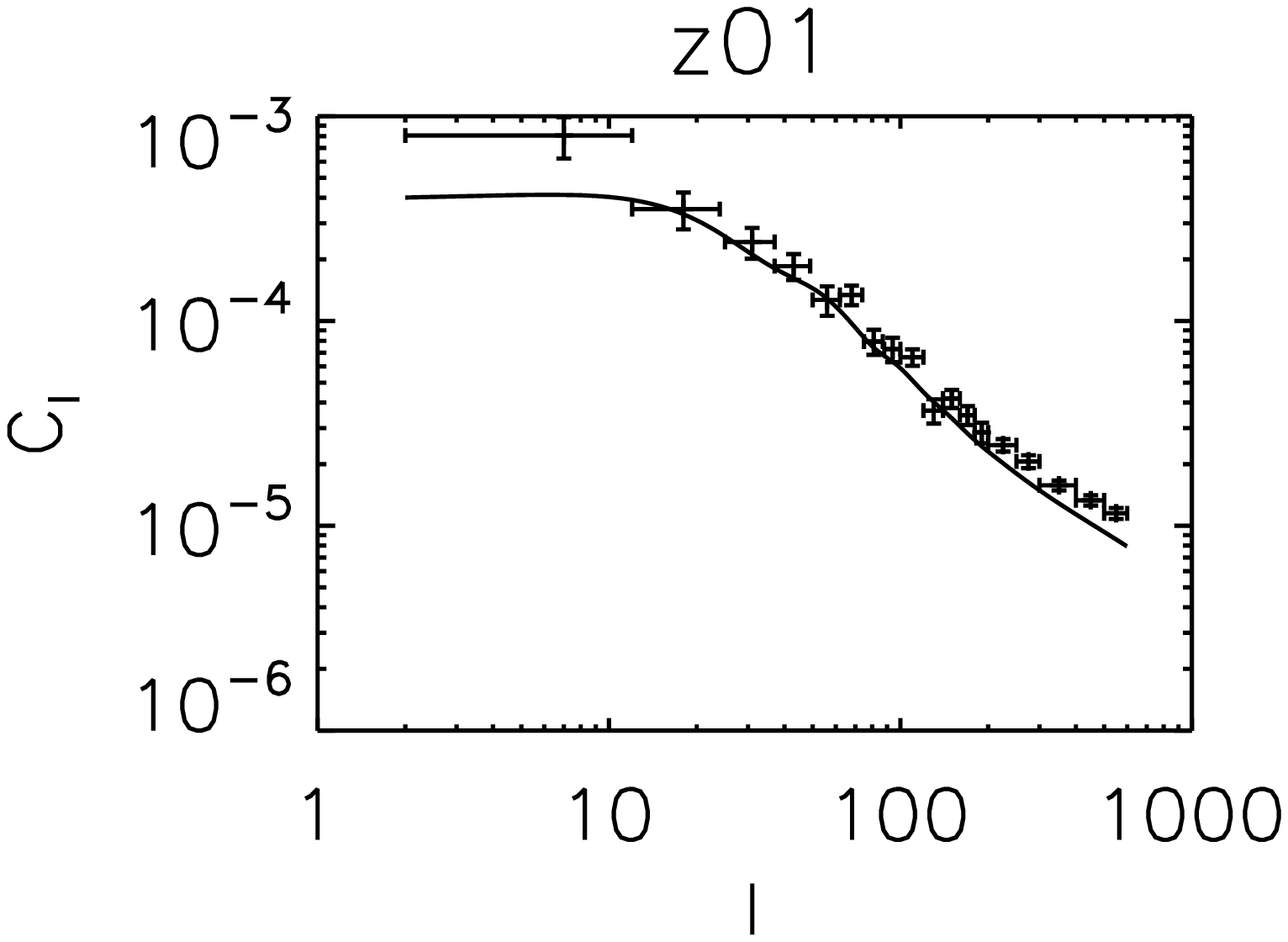}
\includegraphics[width=70 mm]{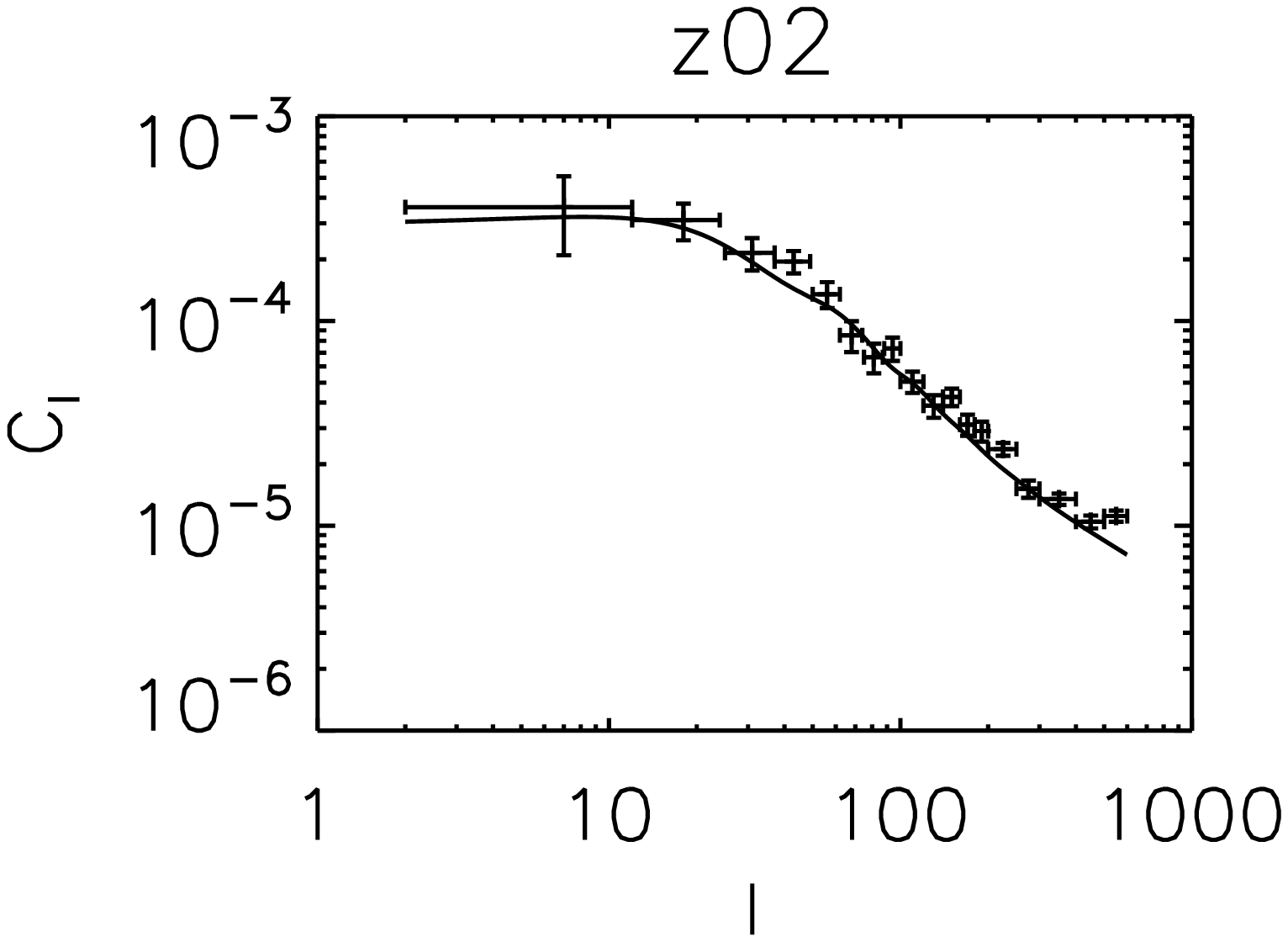}
\includegraphics[width=70 mm]{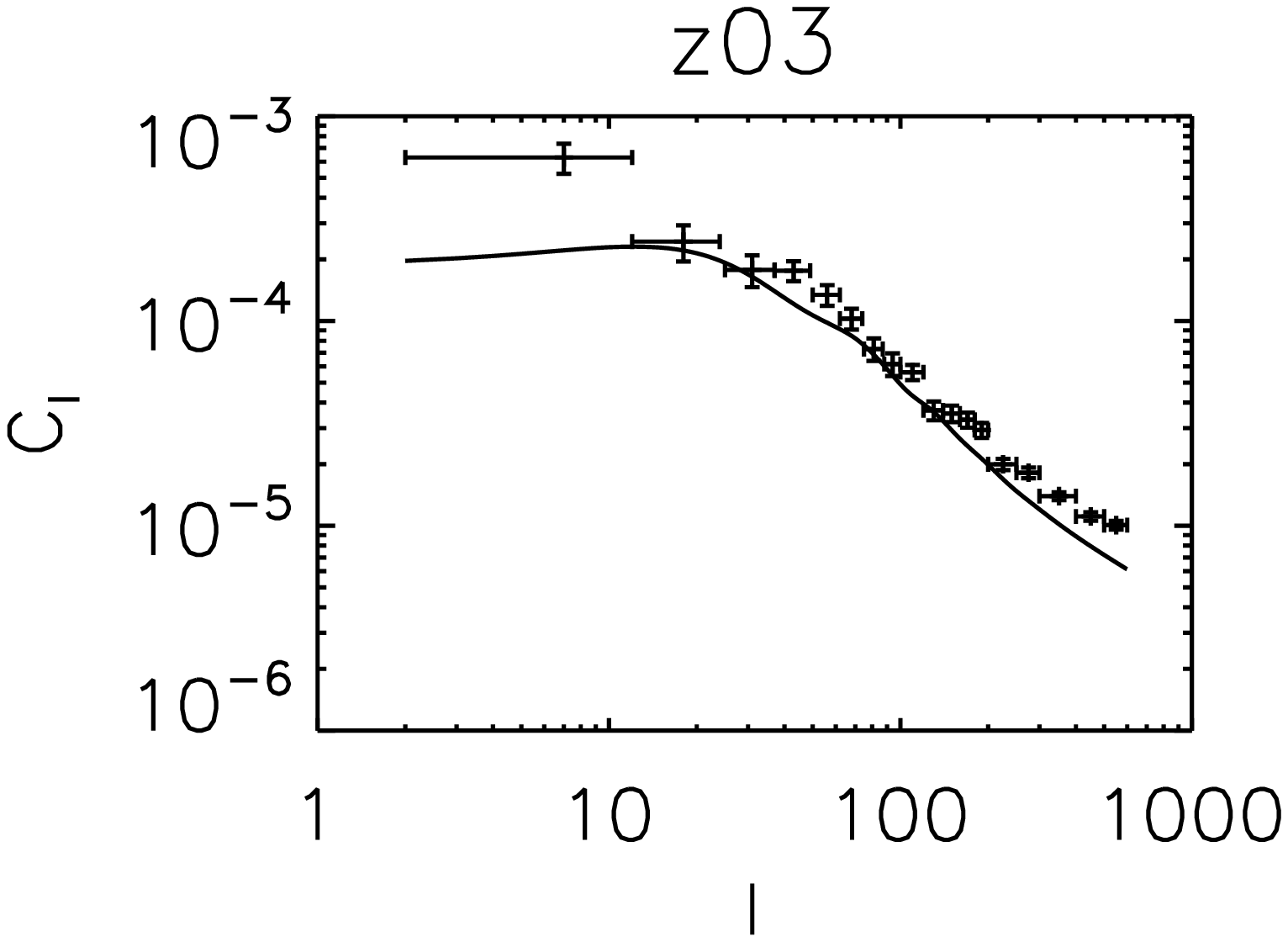}
\includegraphics[width=70 mm]{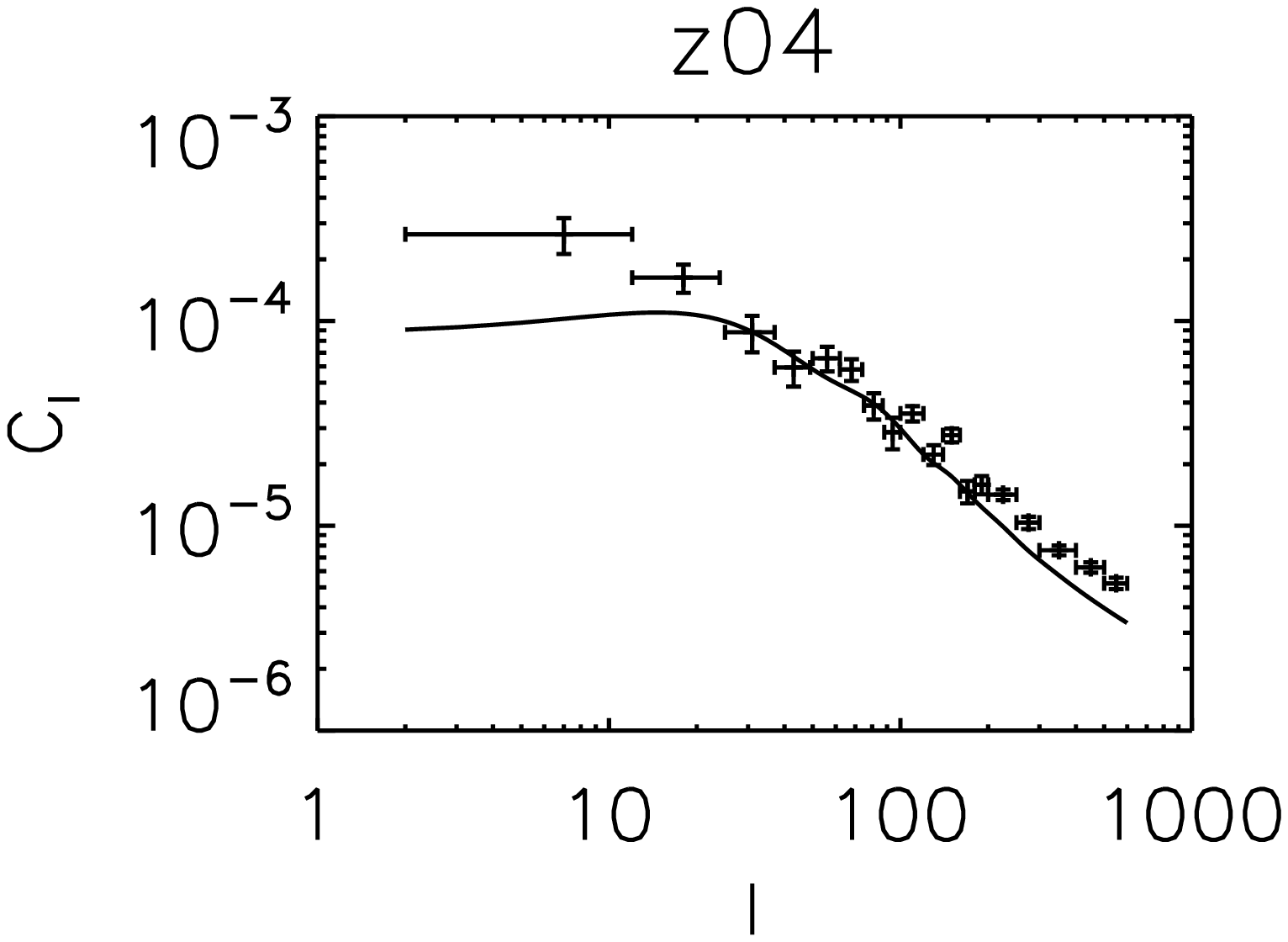}
\includegraphics[width=70 mm]{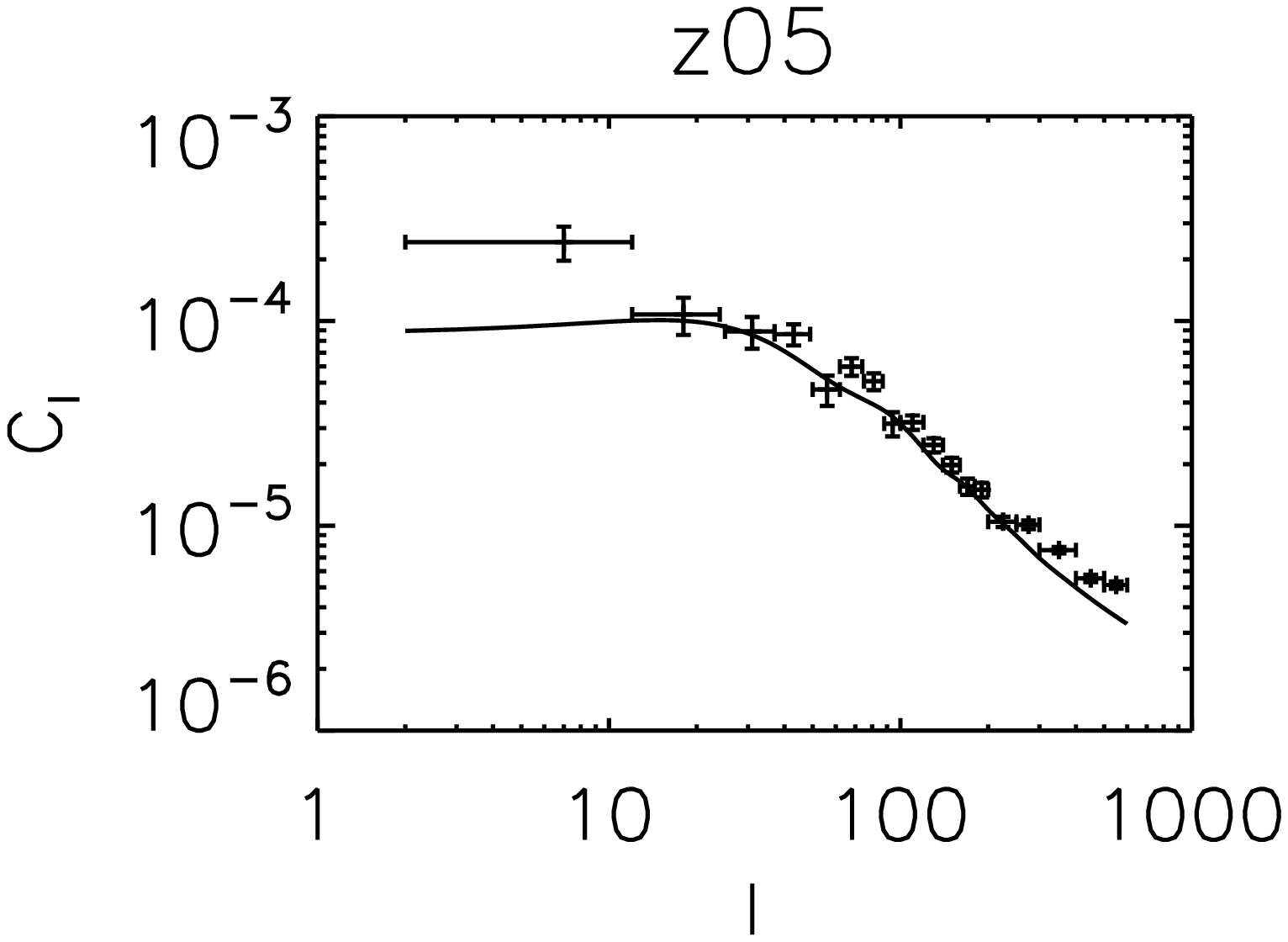}
\includegraphics[width=70 mm]{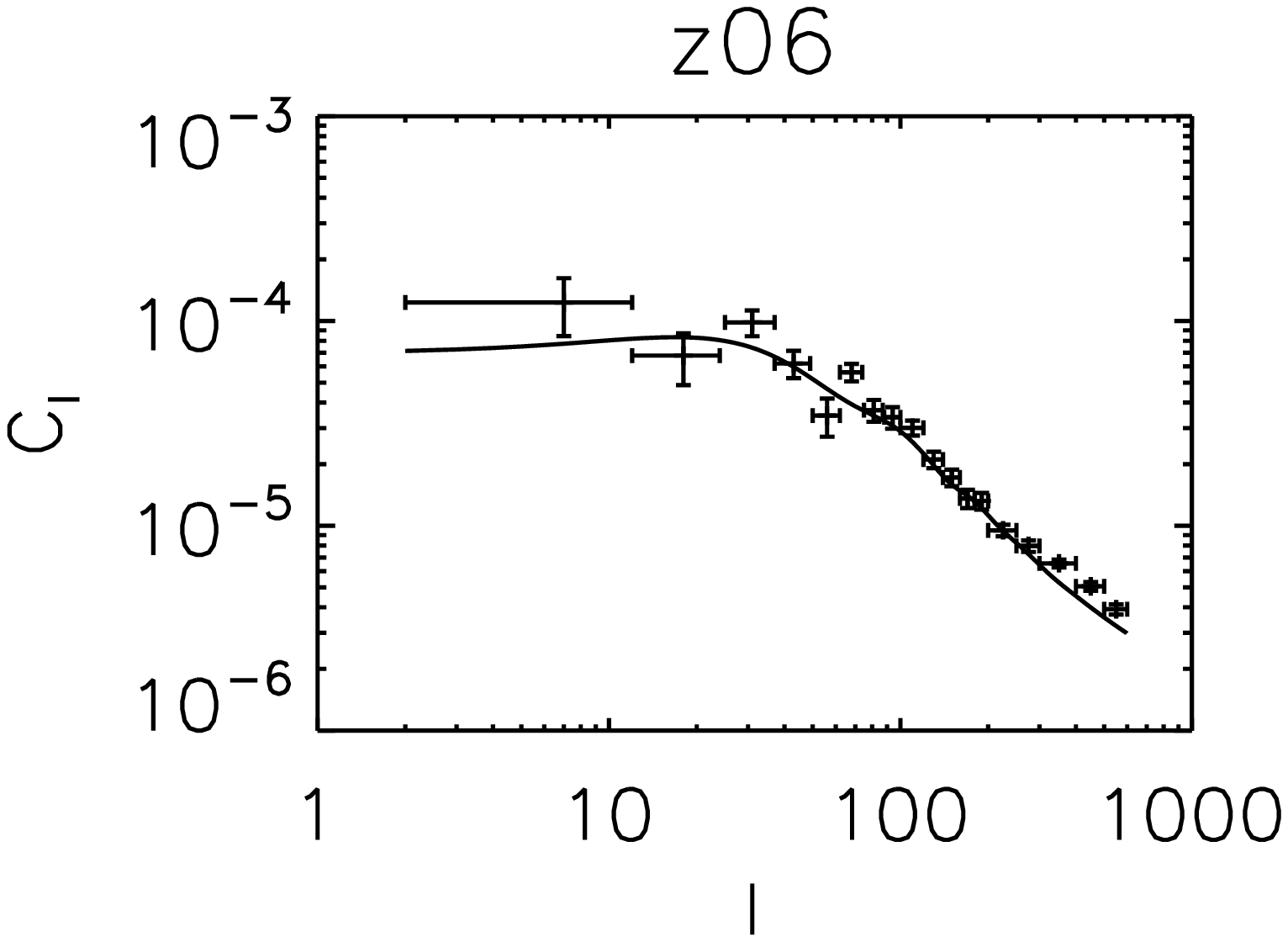}
\includegraphics[width=70 mm]{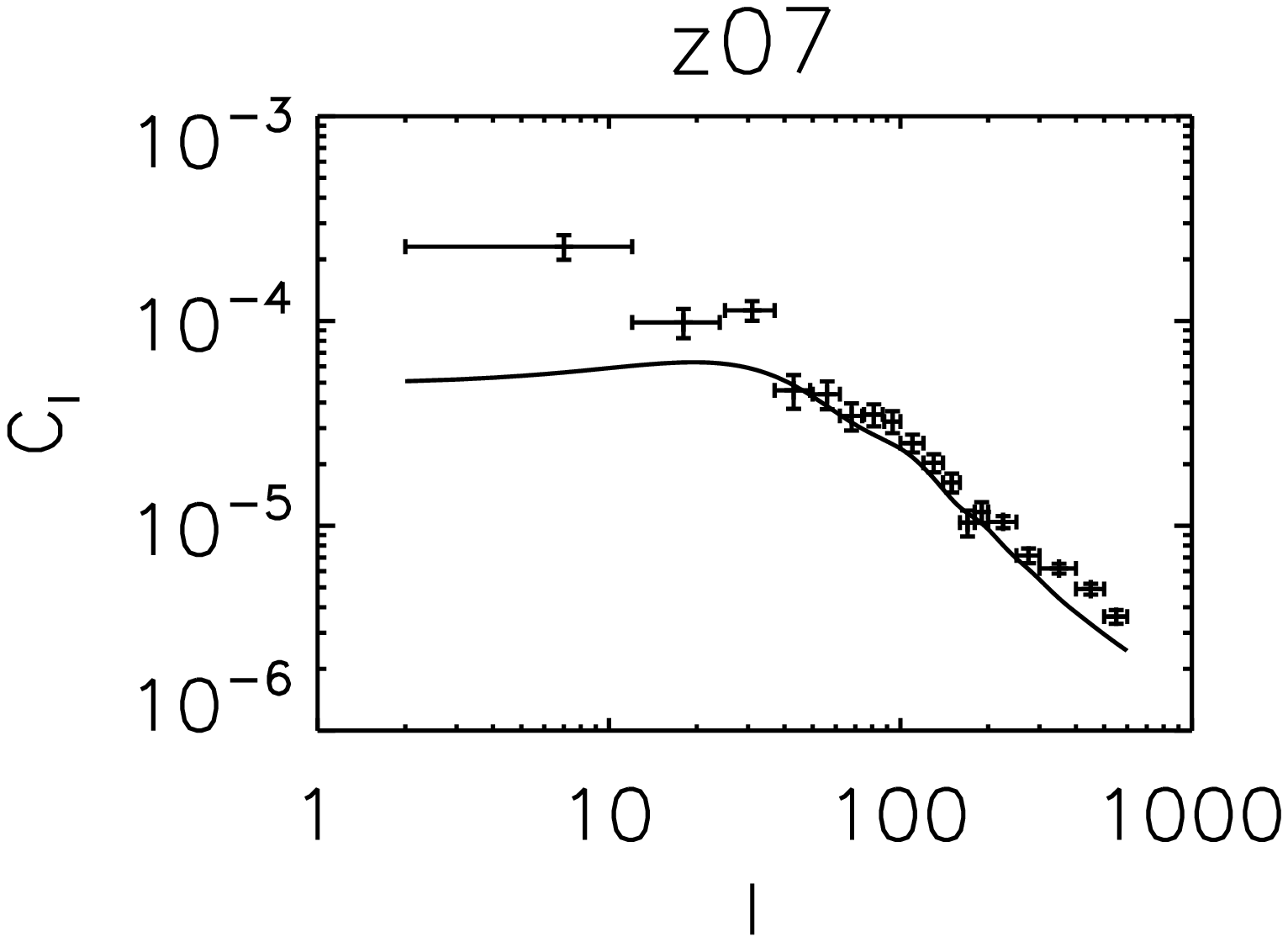}
\caption{The measured angular power spectrum for the eight redshift slices.  The solid lines are the predicted nonlinear power spectra for our fiducial cosmological 
model, 
and the crosses are the measured spectra.\label{F:clresult}}
\end{figure}

We show our results for $C_{g,l}$ in Fig.~\ref{F:clresult}.  Because our galaxy sample covers only 11\% of the sky, the powers in $C_l$ and $g_{2M}$ become 
degenerate, and the isotropic and anisotropic templates become highly correlated.  Thus, when calculating $C_l$, we remove the anisotropic templates for $g_{2M}$, 
$h_{2M}$, and $f_{2M}$ so that power in the $C_l$s is favored.  When using this approach, we see agreement between the measured power spectra and the predicted 
spectra.

\begin{figure}
{\scalebox{.80}{\includegraphics{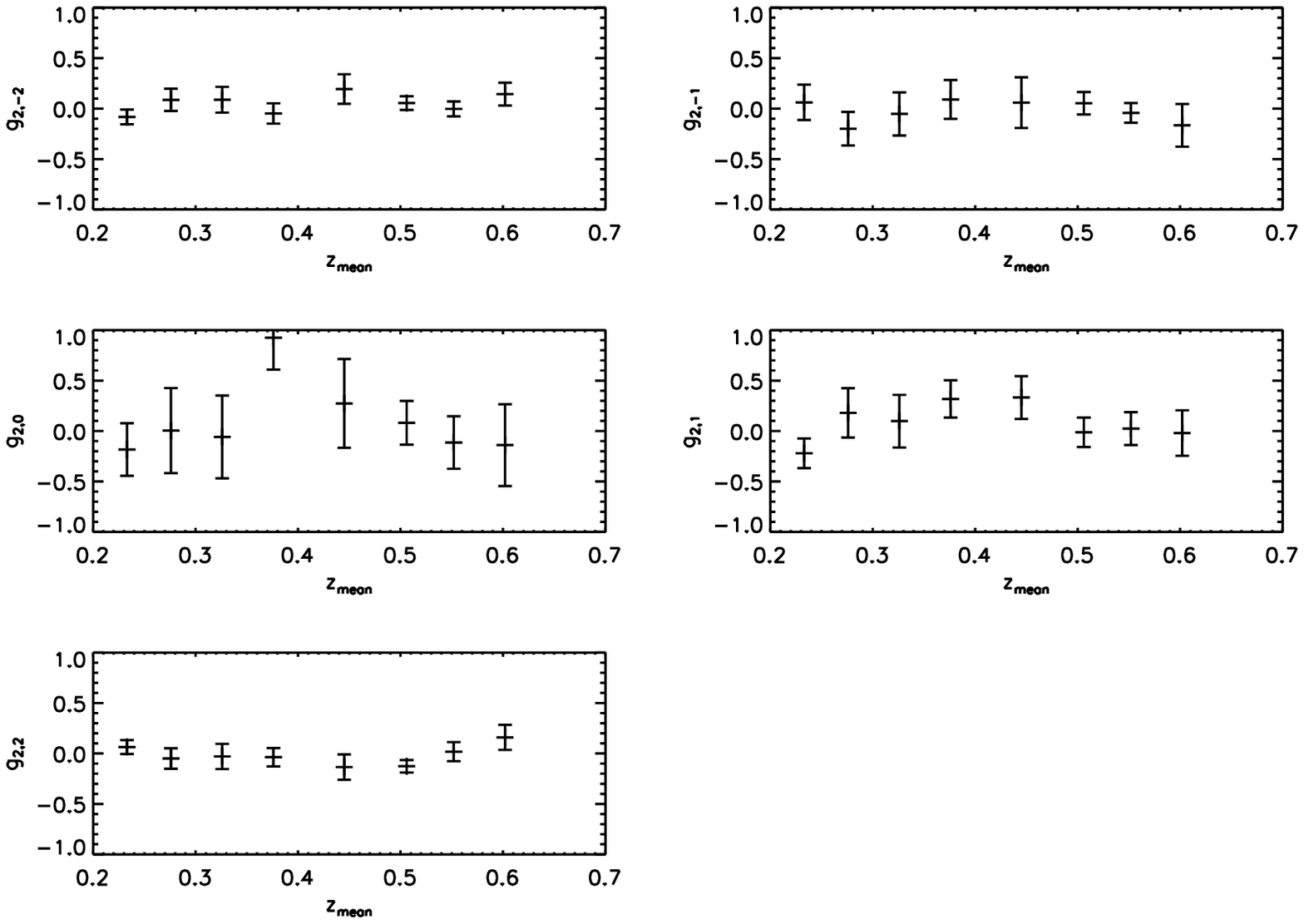}}}
\caption{The quadrupole anisotropy parameters vs.~redshift slice for each multipole with 1$\sigma$ errors from the simulations.  Note $g_{20}=0.925$ for redshift 
slice z03, formally a 
$2.94\sigma$ detection.\label{F:g2mresult}}
\end{figure}

Our results for $g_{2M}$ are shown in Fig.~\ref{F:g2mresult}.  For each multipole of $g_{2M}$, we see consistency with the null result among the redshift slices 
except for the measurement of $g_{20}$ in redshift slice z03.  At this redshift, we measure $g_{20}=0.925\pm0.258$ (Fisher uncertainty) or $\pm 0.315$ (uncertainty 
derived from $N$-body simulations, as described in Sec.~\ref{ss:overall}).  This formally corresponds to a $3.59\sigma$ (Fisher) or $2.94\sigma$ (simulation) 
detection significance; however all of the other redshift slices have $g_{20}$ within $1\sigma$ of zero.  This is puzzling and in principle could indicate either a 
statistical fluke or a systematic error that afflicts the z03 slice.  We note that the statistical significance is marginal: given that we calculated $5\times 8 = 
40$ $g_{2M}$s, the probability of having at least one of them deviate by $2.94\sigma$ is 12\% (assuming a Gaussian distribution).  On the other hand, the z03 slice is also the redshift at which 
the LRG colour locus changes direction \cite{Padmanabhan:2006ku}.  The z03 slice also has the highest bias, which would make is susceptible to nonlinear errors.

\begin{figure}
{\scalebox{.80}{\includegraphics{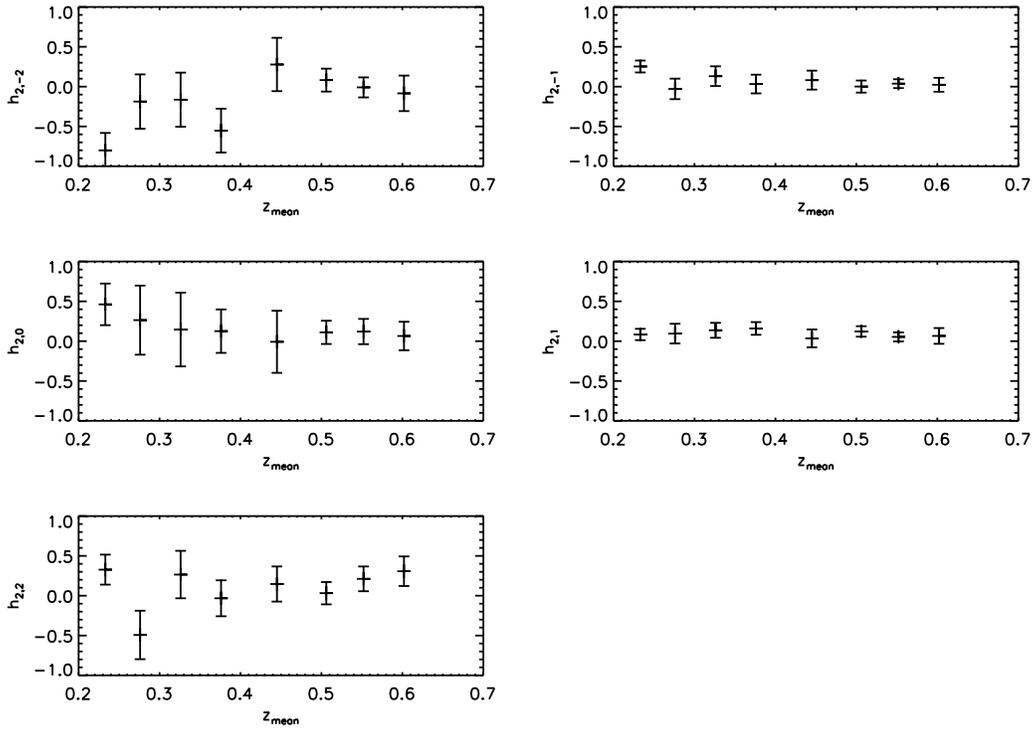}}}
\caption{The quadrupole modulation parameters vs.~redshift slice for each multipole with 1$\sigma$ errors from the simulations.\label{F:h2mresult}}
\end{figure}

\begin{figure}
{\scalebox{.80}{\includegraphics{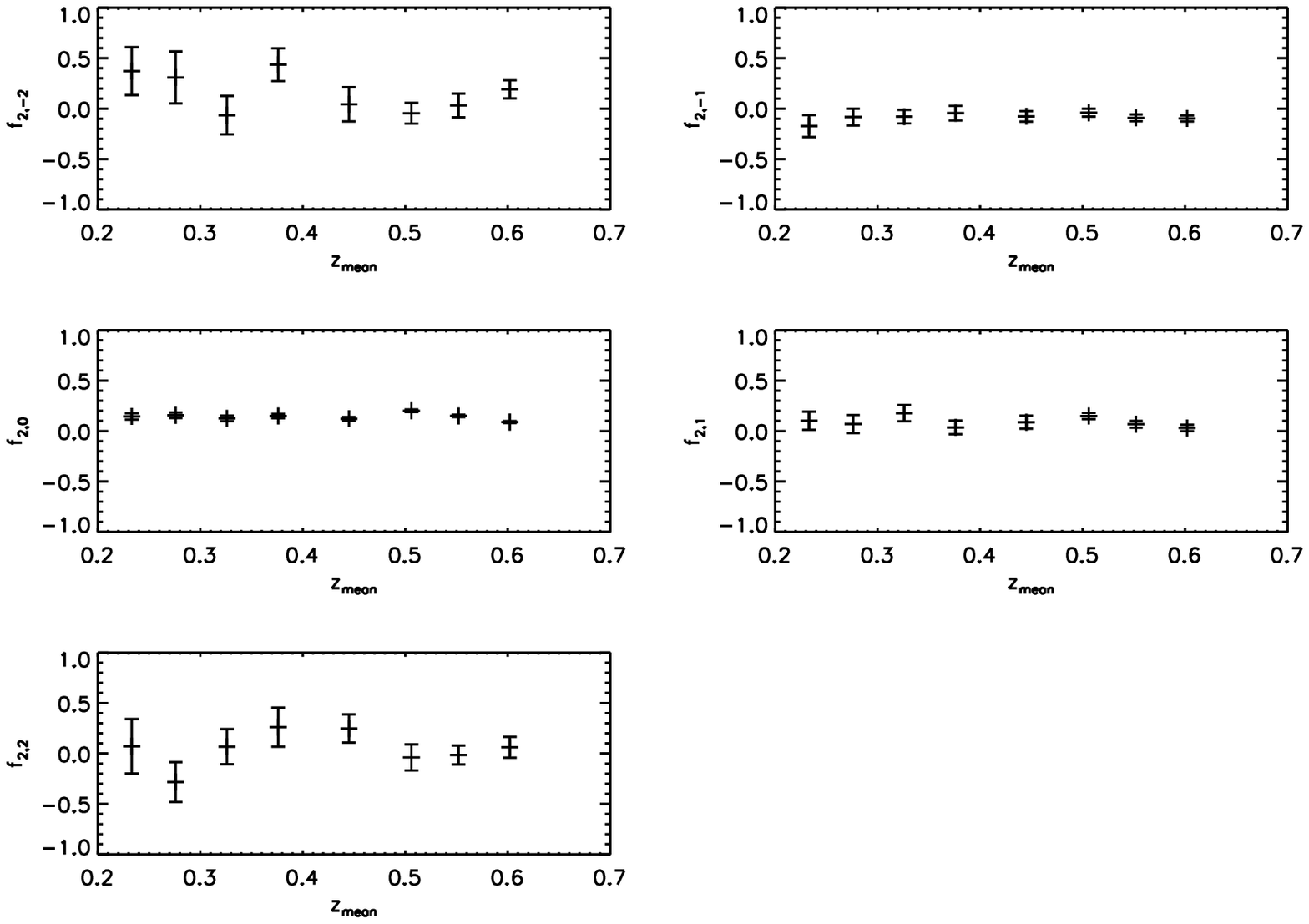}}}
\caption{The Poisson noise modulation parameters vs.~redshift slice for each multipole with 1$\sigma$ errors from the simulations.\label{F:f2mresult}}
\end{figure}

The results we found for the other multipoles were consistent for each redshift slice only when we allowed $h_{2M}$ and $f_{2M}$ to vary from the null result.  We 
show the results for $h_{2M}$ in Fig.~\ref{F:h2mresult} and $f_{2M}$ in Fig.~\ref{F:f2mresult}.  Note that many of the $h_{2M}$s and $f_{2M}$s are inconsistent with 
zero, which hint at possible systematic errors of these forms.

\subsection{Combined statistical anisotropy estimate}
\label{ss:overall}

\begin{figure}
{\scalebox{.80}{\includegraphics{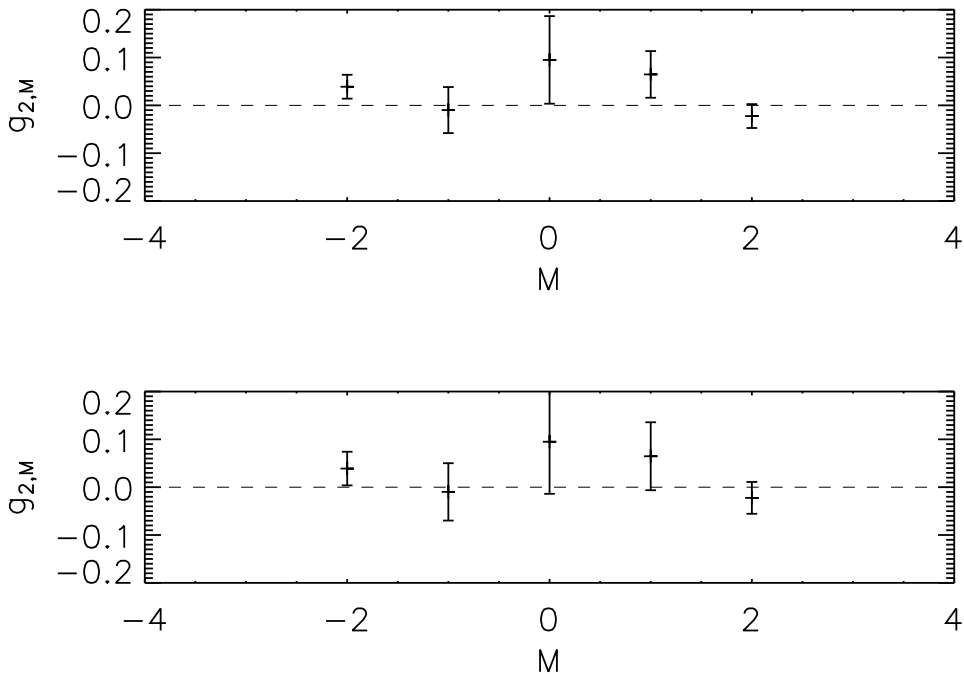}}}
\caption{The quadrupole anisotropy parameters for each multipole marginalized over redshift slice with 1-sigma errors.  The top panel includes errors calculated from 
the 
Fisher matrix. The bottom panel includes errors calculated using N-body simulations.  Note that both results are consistent within two sigma with the null result, shown as the dashed line in both plots.\label{F:g2mf}}
\end{figure}

To find an estimate of $g_{2M}$ combining all of the redshift slices, we construct a minimum-variance estimator of the form
\begin{eqnarray} \label{E:minvar}
\widehat{g}_{2M}=\frac{\sum_i g_{2M,i}/\sigma^2_{g_{2M,i}}}{\sum_i 1/\sigma^2_{g_{2M,i}}}\, ,
\end{eqnarray}
where $g_{2M,i}$ is the estimate of $g_{2M}$ in redshift slice $i$ and $\sigma^2_{g_{2M,i}}=(F^{-1})_{MM}$ for redshift slice $i$.  A crude estimate of the 
uncertainty in $\widehat{g}_{2M}$
is given by
\begin{eqnarray}\label{E:uncvar}
\frac{1}{\sigma^2_{g_{2M}}}=\sum_i\frac{1}{\sigma^2_{g_{2M,i}}}
\end{eqnarray}
This uncertainty estimate neglects covariances between the redshift slices and non-Gaussian (trispectrum) corrections to the errors in individual slices.  We 
therefore expect it to somewhat underestimate the true uncertainty in $\widehat{g}_{2M}$.  For this reason we expect that error bars derived from $N$-body simulations 
(as described next) are more reliable.  The $\widehat{g}_{2M}$ values and their uncertainties as calculated by Eq.~(\ref{E:uncvar}) are shown in the top panel of 
Fig.~\ref{F:g2mf}.

We may alternatively estimate the covariance matrix $C_{MM'}$ of $g_{2M}$ using $N$-body mock catalogues, which contain the correct slice-to-slice correlations and a 
more realistic description of the true non-Gaussian density field.
We used a suite of 
10 simulation boxes of size (2$h^{-1}$ Gpc)$^3$ with periodic boundary conditions, described in more detail in Ref.~\cite{Carlson:2009it}.  For simplicity, and since 
our objective is to obtain a covariance matrix rather than a precision prediction of the power spectrum, we have used the halo catalogue from a single simulation 
output at $z=0.3$.  We populate each halo with a galaxy (or two galaxies if $M_{\rm halo}>M_2$) and use its ``true'' redshift (including the halo peculiar velocity) 
and the photo-$z$ error distribution \cite{Padmanabhan:2004ic} to assign a photometric redshift.  A catalogue of galaxies is then constructed by taking each halo down 
to some minimum mass $M_{{\rm min},i}$ fixed by the requirement to have the correct number of galaxies in the $i$th photo-$z$ slice.  The parameter $M_2$ controls the 
amplitude of the ``1-halo'' term in the power spectrum arising from multiple galaxies per halo (in the sense that the 1-halo term is set to zero if $M_2=\infty$).  
We choose $M_2$ by first constructing a mock catalogue with $M_2=\infty$.  The excess power $\Delta C_l$ in the $300\le l<600$ range is then determined for each 
photo-$z$ slice.  The 3D number density of doubly-occupied haloes $n_d$ required to produce this excess power is then estimated as
\begin{eqnarray}
n_d=\frac{\bar n^2}{2\,dV/d\Omega} \Delta C_l.
\end{eqnarray}
An average value of $n_d$ is taken over all slices ($n_d=1.8\times 10^{-5}h^3\,$Mpc$^{-3}$) and this is used to set a 
mass threshold ($M_2=7.8\times 10^{13}h^{-1}M_\odot$).\footnote{In principle, this procedure could be iterated with computation of a new $\Delta C_l$, etc., however 
the method is probably too crude to justify such a procedure.}  This procedure generates an entire simulated photo-$z$ survey, including all 8 slices and the correct
correlations among different slices due to large scale structure.

We construct 40 realizations of the survey, using each of the 10 boxes 4 times with different observer locations.
We then estimate $g_{2M}$ for each simulation and redshift slice, which we marginalize using Eq.~\ref{E:minvar} to find an estimate of $\widehat{g}_{2M}^{(\alpha)}$ for 
each simulation $\alpha$.  We use these estimates to construct the covariance matrix of the form
\begin{eqnarray} \label{E:covest}
C_{MM'}=\frac{1}{39}\sum_{\alpha=1}^{40}(\widehat{g}_{2M}^{(\alpha)}-\overline{g}_{2M})(\widehat{g}_{2M'}^{(\alpha)}-\overline{g}_{2M'})\, ,
\end{eqnarray}
where $\overline{g}_{2M}$ is $g_{2M}$ averaged over the simulations.  The diagonal elements of $C_{MM'}$ give the uncertainties in $g_{2M}$.  We plot 
$\widehat{g}_{2M}$ with these uncertainties in the bottom panel of Fig.~\ref{F:g2mf}.  We see in this case all the measurements are within one sigma of the null 
result, which is consistent with statistical isotropy.

The final values of $g_{2M}$ and their covariance matrix are given in Table~\ref{T:g2M}.

\begin{table}
\caption{\label{T:g2M}The anisotropy coefficients, averaged over the 8 redshift slices, using the more conservative covariance matrix from the $N$-body simulations.  
Their covariance matrix is also given.}
\begin{indented}
\item[]\begin{tabular}{@{}rrrrrrr}
\br $M$ & $g_{2M}$/10$^{-2}$ & \multicolumn5c{Cov[$g_{2M},g_{2M'}$]/10$^{-3}$} \\
\mr
$-2$ & 3.901 & 1.238 & -0.327 & 0.741 & 0.879 & -0.283 \\
$-1$ & -0.979 & -0.327 & 3.599 & -0.164 & -0.566 & -0.042 \\
$ 0$ & 9.508 & 0.741 & -0.164 & 11.866 & -0.761 & -1.348 \\
$ 1$ & 6.479 & 0.879 & -0.566 & -0.761 & 5.044 & -0.015 \\
$ 2$ & -2.235 & -0.283 & -0.042 & -1.348 & -0.015 & 1.106 \\
\br
\end{tabular}\end{indented}
\end{table}

Finally, as a systematics test, we consider how the $g_{2M}$s change when we do {\em not} project out the extra templates $\{f_{2M},h_{2M}\}$.  We have shown in 
Table~\ref{tab:gfh} the changes in $g_{2M}$ when none of the systematics templates are included (``$g_{2M}$ only'') and when the $f_{2M}$ templates are left out but 
$h_{2M}$ is included (``$g_{2M}$ \& $h_{2M}$'').  As we can see from the table, the exclusion of the $f_{2M}$ templates has essentially no effect, but there is a
substantial change in $g_{2M}$ when the $h_{2M}$ templates are excluded as well.  However, since we expect the main
effect of systematic power spectrum modulation across the sky to be taken into account via the $h_{2M}$s, and given that they change the result by $<3\sigma$,
we do not expect a significant residual systematic after the $h_{2M}$s and $f_{2M}$s have been projected out.

\begin{table}
\caption{\label{tab:gfh}The changes in the anisotropy coefficients, averaged over the 8 redshift slices, when only some of the systematics templates are included.  
We present $\Delta g_{2M}$ values, which are equal to the $g_{2M}$ from the full analysis minus those with only some of the systematics templates, and the 
number of sigmas by which the correction differs from zero, $\Delta g_{2M}/\sigma(\Delta g_{2M})$.}
\begin{indented}
\item[]\begin{tabular}{@{}rrrr}
\br & \multicolumn2c{$g_{2M}$ only} & $g_{2M}$ \& $h_{2M}$ \\
$M$ & $\Delta g_{2M}$ & $\Delta g_{2M}/\sigma(g_{2M})$ & $\Delta g_{2M}$ \\
\mr
$-2$ & $ 0.0128$ & $0.63$ & $ 0.0004$ \\
$-1$ & $ 0.1591$ & $2.02$ & $ 0.0027$ \\
$ 0$ & $ 0.2827$ & $1.94$ & $-0.0050$ \\
$ 1$ & $ 0.2830$ & $2.78$ & $ 0.0004$ \\
$ 2$ & $-0.0003$ & $0.02$ & $-0.0003$ \\
\br
\end{tabular}\end{indented}
\end{table}

\subsection{Comparison with CMB results}
\label{s:cmb}

Groeneboom {\em et~al.} \cite{G09} report evidence for a quadrupolar power asymmetry in the 5-year WMAP data.  They investigated models of the form:
\begin{equation}
P(\veck) = P(k)[1 + g_*(\hatk\cdot\hatn)^2],
\end{equation}
where $g_*$ is the amplitude of the asymmetry and $\hatn$ is its preferred axis.  For $|g_*|\ll1$, this is equivalent to our Eq.~(\ref{E:powerspectrum}) with
\begin{equation}
\sum_{LM} g_{LM} R_{LM}(\hatk) = g_*\left[ (\hatk\cdot\hatn)^2 - \frac13 \right] = \frac23g_* P_2(\hatk\cdot\hatn)
\label{E:ps2}
\end{equation}
and a slightly rescaled definition of the power spectrum, $P(k)=\bar P(k)(1-\frac13g_*)$.  Here the $-\frac13$ ensures that there is no $L=0$ term in Eq.~(\ref{E:ps2}).  
Using the spherical harmonic addition theorem, we can see that this requires
\begin{equation}
g_{LM} = \frac{8\pi}{15}g_* \delta_{L2} R_{2M}(\hatn).
\label{E:g2}
\end{equation}
We may use this to construct an estimator for $g_*$ assuming a particular direction $\hatn$; here we will take  $\hatn$ to be in the Groeneboom \& Eriksen direction 
so that we can test for consistency with their value of $g_*$.  The best estimator is
\begin{equation}
\hat g_* = \frac{15}{8\pi} \frac{ \sum_{MM'} [{\rm Cov}^{-1}]_{MM'} \hat g_{2M} R_{2M'}(\hatn) }{ \sum_{MM'} [{\rm Cov}^{-1}]_{MM'} R_{2M}(\hatn) R_{2M'}(\hatn) },
\end{equation}
with uncertainty
\begin{equation}
\sigma(\hat g_*) = \frac{15}{8\pi} \frac{1}{\sqrt{ \sum_{MM'} [{\rm Cov}^{-1}]_{MM'} R_{2M}(\hatn) 
R_{2M'}(\hatn) }};
\end{equation}
here Cov is the $5\times5$ covariance matrix of the estimators for $g_{2M}$ (see Table~\ref{T:g2M}).

Using the WMAP W-band maps, and considering multipoles in the CMB out to $l_{\rm max}=400$, Groeneboom {\em et~al.}~\cite{G09} find an asymmetry of $g_*^{\rm CMB} = 
0.29\pm0.031$ with an axis of maximum power in the direction $(l,b)=(94^\circ,26^\circ)$, which they attribute to an unknown systematic effect because different 
signals are observed in the V and W bands and the apparent alignment with the Ecliptic Poles.  Using the above projection procedure, we find an amplitude $g_*^{\rm 
LRG}=0.006\pm0.036$ in this direction.  Groeneboom {\em et~al.} also did their fit using the WMAP V-band maps, finding $g_*^{\rm CMB}=0.14\pm0.034$ in the direction 
$(l,b)=(97^\circ,27^\circ)$; when we project our LRG anisotropy coefficients onto this axis, we find $g_*^{\rm LRG}=0.007\pm0.037$.  Foregrounds and noise 
mis-estimation have been disfavored as possible candidates \cite{Bennett:2010jb}.  A possible cause for the appearance of statistical anisotropy in the CMB data would 
be the ellipticity of the WMAP beams, which when combined with the survey strategy could result in a preferred axis in the direction of the Ecliptic Poles 
\cite{Groeneboom:2008fz,Wehus:2009zh,Hanson:2009gu,Hanson:2010gu}.  Specifically, Hanson {\em et al.}~\cite{Hanson:2010gu} find that once asymmetric beam effects are 
subtracted, the data is consistent with the isotropic model; however Groeneboom {\em et~al.}~\cite{G09} evaluated the resulting effect and found it to be negligible.  
The WMAP 7-year analysis finds no known instrumental effect other than beam asymmetry that can cause the anomaly, but they have not yet completed a full simulation of 
beam asymmetry effects on quadrupolar power modulation \cite{Bennett:2010jb}.  We also note the the WMAP team has already accounted for these beam effects in their 
estimation of the power spectrum, so this systematic in the quadrupolar anisotropy does {\em not} affect the cosmological parameters derived from WMAP.

Thus the cause of the apparent asymmetry in the WMAP maps is not definitively known.  In any case, our LRG analysis finds no anisotropy in this direction.  It is 
possible that $g_{2M}$ is different at the two scales probed by the CMB and the LRG sample.  In most variants of inflation, where each $e$-fold of expansion is 
similar to the previous one with $\sim {\cal O}(1/N)$ deviations (where $N$ is the number of $e$-folds remaining), we would expect $g_{2M}$ to
vary smoothly with the number of $e$-folds, or $\ln k$.  We find the effective scales probed by Groeneboom {\em et al.}'s CMB analysis \cite{G09} and our LRG analysis 
are 0.020 Mpc$^{-1}$ and 0.15 Mpc$^{-1}$, respectively (see \ref{A:effscale}), which differ by only 2.0 $e$-folds.  It would be very surprising if inflation were not 
only anisotropic but also managed to produce a scale-dependent anisotropy that varied over so short a baseline.

\subsection{Direction-marginalized constraint on $g_*$}
\label{ss:dir}

The above analyses have either set constraints on a general $g_{2M}$ (a 5-dimensional parameter space) or on $g_*$ for a fixed anisotropy axis (a 1-dimensional parameter space).  It is however of 
interest to set constraints on general axisymmetric quadrupolar anisotropies, such as Eq.~(\ref{E:ps2}), which would arise if there were a single preferred axis during inflation.  This is a 3-dimensional 
parameter space: there is an amplitude $g_*$ and a direction $\hatn\in$S$^2$.

We may set constraints on $g_*$ via a Bayesian analysis in which a uniform prior is placed on $\hatn$, as has been done in several previous statistical anisotropy analyses \cite{Groeneboom:2008fz, 
Hirata:2009ar}.  Our problem -- setting a limit on the amplitude of an anisotropy while marginalizing over its direction -- is similar to that performed by Ref.~\cite{Hirata:2009ar} for the large scale 
structure dipole; we follow the same methodology, although we note that for the quadrupolar asymmetry $g_*$ could be positive or negative (``prolate'' and ``oblate'' power anisotropies are different and 
cannot be rotated into each other).  The marginalized likelihood function for $g_*$ is
\begin{eqnarray}
{\cal L}(g_*) &=& \int \exp\Biggl\{-\frac12\sum_{MM'} [{\rm Cov}^{-1}]_{MM'}
\left[ \hat g_{2M} - \frac{8\pi}{15}g_*R_{2M}(\hatn) \right]
\nonumber \\ && \times
\left[ \hat g_{2M'} - \frac{8\pi}{15}g_*R_{2M'}(\hatn) \right]
\Biggr\}\,d^2\hatn,
\end{eqnarray}
where $\hat g_{2M}$ are the estimated anisotropy coefficients and Cov is their $5\times 5$ covariance matrix.  If we set a uniform prior on $g_*$, as done by Groeneboom \& Eriksen 
\cite{Groeneboom:2008fz}, then we may divide ${\cal L}(g_*)$ by its integral $\int {\cal L}(g_*)\,dg_*$ and treat it as a posterior probability distribution.  If we do this, then we find that 68\% of the 
posterior distribution is contained within $-0.12<g_*<+0.10$ and 95\% within $-0.41<g_*<+0.38$.  Note that the distribution has very non-Gaussian tails because of the large uncertainty on $g_{20}$: a 
quadrupole anisotropy aligned with the Galactic axis would be difficult to detect given our sky coverage.  There is a small probability for such an alignment to occur and not produce measurable $g_{2M}$ 
($M\neq0$) even if $g_*$ is large.

\section{Conclusions} \label{S:conclude}

We have conducted a search for statistical anisotropy in the galaxy distribution.  Statistical anisotropy can manifest from the direction-dependent primordial power 
spectrum shown in Eq.~\ref{E:powerspectrum} with the magnitude of the anisotropy parametrized by $g_{LM}$.  This phenomenon causes the angular galaxy power spectrum 
$C_{g,l}$ to be generalized by $D_{g,ll'}^{LM}$, which includes $g_{LM}$.  We used estimators formulated by Padmanabhan {\em et al.}~\cite{Padmanabhan:2006ku} and a sample 
of LRGs from SDSS to search for evidence of quadrupolar anisotropy parametrized by $g_{2M}$.  We found $g_{2M}$ for all $M$ to be within $2\sigma$ of zero.  Using our 
estimates of $g_{2M}$ and assuming a symmetry axis in the direction $(l,b)=(94^\circ,26^\circ)$, we calculated the anisotropy amplitude 
$g_*=0.006\pm0.036\,(1\sigma)$.  This confirms that the previously identified anisotropy in the WMAP maps (already believed to be a systematic effect) is not of 
primordial origin.
When marginalizing over the symmetry axis direction and assuming a uniform prior for $g_*$, we constrain $-0.41<g_*<+0.38$ with a 
95\% confidence level.

Looking forward, we expect much better sensitivity to $g_*$ from future galaxy surveys.  For fixed sky coverage, the uncertainty in $g_{2M}$ is proportional to the 
inverse square-root of the number of modes measured, i.e. it is proportional to $l_{\rm max}^{-1}N_z^{-1/2}$ where $l_{\rm max}$ is the maximum multipole at which the 
galaxy distribution is well-sampled, and $N_z$ is the number of effectively independent redshift slices.  The largest advance may be possible with future large-volume 
spectroscopic surveys intended to study baryon oscillations.  Here the effective number of redshift slices is $N_z\sim k_{\rm max}\Delta r/\pi$, where $\Delta r$ is 
the radial width of the survey; for surveys that reach out to $z\approx 2$ this is $N_z\sim 100$ (instead of 8 here).  As this redshift corresponds to a factor of 
$\sim 3$ increase in distance relative to the SDSS LRGs, we would expect that for similar sampling $nP(k)$ $l_{\rm max}$ should increase by a factor of 3.  Thus such 
a survey should in principle be able to improve measurements of $g_{2M}$ by an order of magnitude relative to those presented here.  Further improvements in $g_{20}$ 
might also be possible if improvements in the dust map or work in redder bands allows one to work at lower Galactic latitudes.

\ack

We thank N Padmanabhan and S Ho for providing the LRG sample and for their useful feedback.  We also thank M Kamionkowski for helpful comments.

AP acknowledges the support of the NSF.  This work was supported by DOE DE-FG03-92-ER40701, NASA NNG05GF69G, the Gordon and Betty Moore Foundation, and a NASA 
Einstein Probe mission study grant, ``The Experimental Probe of Inflationary Cosmology."

CH is supported by the US Department of Energy under contract DE-FG03-02-ER40701, the National Science Foundation under contract AST-0807337, and the
Alfred P Sloan Foundation.

Funding for the Sloan Digital Sky Survey (SDSS) and SDSS-II has been provided by the Alfred P Sloan Foundation, the Participating Institutions,
the National Science Foundation, the US Department of Energy, the National Aeronautics and Space Administration, the Japanese Monbukagakusho, and the
Max Planck Society, and the Higher Education Funding Council for England. The SDSS Web site is http://www.sdss.org/.

The SDSS is managed by the Astrophysical Research Consortium (ARC) for the Participating Institutions. The Participating Institutions are the
American Museum of Natural History, Astrophysical Institute Potsdam, University of Basel, University of Cambridge, Case Western Reserve University, The
University of Chicago, Drexel University, Fermilab, the Institute for Advanced Study, the Japan Participation Group, The Johns Hopkins University, the
Joint Institute for Nuclear Astrophysics, the Kavli Institute for Particle Astrophysics and Cosmology, the Korean Scientist Group, the Chinese Academy
of Sciences (LAMOST), Los Alamos National Laboratory, the Max-Planck-Institute for Astronomy (MPIA), the Max-Planck-Institute for Astrophysics (MPA),
New Mexico State University, Ohio State University, University of Pittsburgh, University of Portsmouth, Princeton University, the United States Naval
Observatory, and the University of Washington.

\appendix

\section{Real $l=2$ spherical harmonics} \label{A:realhar}

In Eq.~\ref{E:realcomplex} we introduce our convention for the real spherical harmonics $R_{LM}(\theta,\phi)$.  To clarify the functional form of $R_{LM}$,  we list the harmonics for $L=2$.  These are given by
\begin{eqnarray} \label{E:r2}
R_{22}(\theta,\phi) &=& \sqrt{\frac{15}{16\pi}}\sin^2\theta\cos(2\phi) \nonumber \\
R_{21}(\theta,\phi) &=& -\sqrt{\frac{15}{4\pi}}\cos\theta\sin\theta\cos\phi \nonumber \\
R_{20}(\theta,\phi) &=& \sqrt{\frac{5}{16\pi}}\left(3\cos^2\theta-1\right) \nonumber \\
R_{2,-1}(\theta,\phi) &=& \sqrt{\frac{15}{4\pi}}\cos\theta\sin\theta\sin\phi \nonumber \\
R_{2,-2}(\theta,\phi) &=& -\sqrt{\frac{15}{16\pi}}\sin^2\theta\sin(2\phi) \, .
\end{eqnarray}

\section{Expressions for the anisotropy coefficient} \label{A:anicoef}

In Ref.~\cite{Pullen:2007tu}, Pullen and Kamionkowski introduced an anisotropy coefficient $\xi_{lml'm'}^{LM}$ that appears in the correlation function, given by
\begin{eqnarray} \label{E:xint}
     \xi^{LM}_{lml^\prime m^\prime} &=&
     \int d\mathbf{\hat{k}} \, Y_{lm}^\ast(\mathbf{\hat{k}})
     Y_{l^\prime m^\prime}(\mathbf{\hat{k}})
     Y_{LM}(\mathbf{\hat{k}}) \nonumber \\
     &=& (-1)^m \left(G^L_{ll'} \right)^{1/2}
     C^{LM}_{lml',-m'}\, ,
\end{eqnarray}
where $C_{lml'm'}^{LM}$ are Clebsch-Gordan coefficients, and
\begin{equation}
     G^L_{ll'} \equiv \frac{ (2l+1)(2l'+1)}{ 4 \pi (2L+1) }
     \left(C^{L0}_{l0l'0} \right)^2.
\end{equation}
However, since we use real spherical harmonics (given by Eq.~\ref{E:realcomplex}) in our analysis as opposed to complex spherical harmonics, we introduce the 
anisotropy coefficient $X_{lml'm'}^{LM}$ given by Eq.~(\ref{E:xlm}).  We choose to write $X_{lml'm'}^{LM}$ in terms of Wigner 3j symbols.  Due to the piecewise nature 
of the real spherical harmonics, $X_{lml'm'}^{LM}$ will have different expressions for different values of $m$, $m'$, and $M$.  After much algebra, we can find the 
expressions for $X_{lml'm'}^{LM}$ in terms of Wigner 3j symbols (written in matrix form) and $P_{ll'L}$, given by
\begin{eqnarray} \label{E:pL}
P_{ll'L}=\sqrt{\frac{(2l+1)(2l'+1)(2L+1)}{4\pi}}\left(\begin{array}{ccc}l&l'&L\\0&0&0\end{array}\right),
\end{eqnarray}
which is nonzero only for $l+l'+L$ even.  The expression for $M=0$ is given by
\begin{eqnarray} \label{E:xlm0}
X_{lml'm'}^{L0}=(-1)^mP_{ll'L}\left(\begin{array}{ccc}l&l'&L\\m&-m&0\end{array}\right)\delta_{mm'}.
\end{eqnarray}
The expressions for $M\neq 0$ can be obtained similarly, e.g.
\begin{eqnarray}
X_{lml'm'}^{LM}&=&P_{ll'L}\left[\frac{(-1)^{m'}}{\sqrt{2}}\left(\begin{array}{ccc}l&l'&L\\m&-(m+M)&M\end{array}\right)\delta_{m',m+M}\right.\nonumber\\
&&\left.+\frac{(-1)^m}{\sqrt{2}}\left(\begin{array}{ccc}l&l'&L\\m&M-m&-M\end{array}\right)\delta_{m',m-M}\right];
\end{eqnarray}
for $m>M>0$; the other equations are similar but will be omitted for brevity.

\section{Effective scale for quadrupole asymmetry analyses} \label{A:effscale}

In Ref.~\cite{G09}, Groeneboom et.~al.~calculated a quadrupole asymmetry in the matter power spectrum by analysis of the CMB up to multipoles of $l=400$.  In this appendix, we derive the effective wavenumber $k_{\rm{eff}}$ of this CMB measurement, as well as $k_{\rm{eff}}$ for our measurement using the LRG sample.

To find $k_{\rm{eff}}$ for the CMB analysis, we first find an estimator for $g_{2M}$ in terms of measurable quantities in Fourier space.\footnote{Although Ref.~\cite{G09} parametrized the quadrupole asymmetry in terms of $g_*$, not $g_{2M}$, this should not affect the effective wavenumber of the measurement.}  This calculation has been done previously in Ref.~\cite{Pullen:2007tu}.  By using Eq. (23) in that paper, we can construct the minimum-variance estimator for $g_{2M}$ given by
\begin{eqnarray} \label{E:estg}
\widehat{g}_{2M}=\frac{\sum_{ll'}\widehat{g}_{2M,ll'}Q_{ll'2}(F_{ll'}^{\rm{map}})^2/(C_l^{\rm{map}}C_{l'}^{\rm{map}})}{\sum_{ll'}Q_{ll'2}(F_{ll'}^{\rm{map}})^2/(C_l^{\rm{map}}C_{l'}^{\rm{map}})}\, ,
\end{eqnarray}
where $C_l^{\rm{map}}=|W_l|^2C_l+C_l^n$ is the map's power spectrum, $W_l=e^{-l^2\sigma_b^2/2}$ is the beam window function, $F_{ll'}^{\rm{map}}=W_lW_{l'}F_{ll'}$, $Q_{ll'2}=\sum_{mm'}(X_{lml'm'}^{2M})^2=(P_{ll'2})^2/5$, and $\widehat{g}_{2M,ll'}$ is the estimator of $g_{2M}$.  We can construct an estimator for each $ll'$ pair with $\widehat{g}_{2M,ll'}=\widehat{D}_{ll'}^{2M}/F_{ll'}$, where $\widehat{D}_{ll'}^{2M}$ is an estimator constructed from the measured $a_{lm}$s.  By approximating for large $l$
\begin{eqnarray} \label{E:3japp}
\left(\begin{array}{ccc}l&l&2\\0&0&0\end{array}\right)^2\simeq\frac{1}{8l} \mbox{ and} \left(\begin{array}{ccc}l&l\pm2&2\\0&0&0\end{array}\right)^2\simeq\frac{3}{16l}\, ,
\end{eqnarray}
using Eq. (8.32) of Ref.~\cite{Varshalovich:1988ye}, along with $F_{ll}=C_l$ and $F_{l(l\pm2)}\simeq-0.5C_l$ (for temperature perturbations), we have
\begin{eqnarray} \label{E:estgapp}
\widehat{g}_{2M}\simeq\frac{\sum_l lT_l[\frac{1}{8\pi}\widehat{g}_{2M,ll}+\frac{3}{64\pi}(\widehat{g}_{2M,l(l+2)}+\widehat{g}_{2M,l(l-2)})]}{\frac{7}{32\pi}\sum_l lT_l} \, ,
\end{eqnarray}
where $T_l=[1+C_l^n/(W_l^2C_l)]^{-2}$.

If the expectation value of this estimator is taken, then $\widehat{g}_{2M,ll'}$ is replaced with $\VEV{\widehat{g}_{2M,ll'}}=D_{ll'}^{2M}/F_{ll'}$.  This estimator was constructed with the assumption that $g_{2M}$ is scale-invariant, in which case $\VEV{\widehat{g}_{2M,ll'}}=g_{2M}$ and $\VEV{\widehat{g}_{2M}}$ does give the true $g_{2M}$.  However, if $g_{2M}$ does vary with scale, then it has to be included inside the integral for $D_{ll'}^{2M}$, and the estimator's expectation value must be taken as $g_{2M}$ at $k=k_{\rm{eff}}$.  We find this scale by giving $g_{2M}(k)$ a functional form, which we use to find $\VEV{\widehat{g}_{2M}}=g_{2M}(k_{\rm{eff}})$.

We begin with the expectation value of $\widehat{g}_{2M,ll'}$, given by
\begin{eqnarray}\label{E:estgl}
\VEV{\widehat{g}_{2M,ll'}} = \frac{\int g_{2M}(k)\bar P_\zeta(k) \Theta_l(k)\Theta_{l'}(k)k^2dk}{\int \bar P_\zeta(k) \Theta_l(k)\Theta_{l'}(k)k^2dk}\, ,
\end{eqnarray}
where $\bar P_\zeta$ is the curvature power spectrum and $\Theta_l(k)$ is the transfer function of the CMB temperature fluctuations.  Typically, in inflationary models that break scale invariance, $g_{2M}$ will vary smoothly with $\ln k$, and we can write
\begin{eqnarray} \label{E:gvar}
g_{2M}(k) = B_1+B_2\ln (k/k_*)\, ,
\end{eqnarray}
where $k_*$ is the (arbitrary) pivot wavenumber such that $\bar P_\zeta(k_*)$ is constant even when the scalar spectral index $n_s$ is varied.  We then insert this expression into Eq.~\ref{E:estgl} and find
\begin{eqnarray} \label{E:estglb}
\VEV{\widehat{g}_{2M,ll'}}= B_1+B_2\frac{\int \ln(k/k_*)\bar P_\zeta(k) \Theta_l(k)\Theta_{l'}(k)k^2dk}{\int \bar P_\zeta(k) \Theta_l(k)\Theta_{l'}(k)k^2dk}\, .
\end{eqnarray}
The denominator in the second term is just $F_{ll'}$.  The numerator can be rewritten by using $\bar P_\zeta(k)\propto (k/k_*)^{n_s}$ such that
\begin{eqnarray} \label{D:estgibd}
 \VEV{\widehat{g}_{2M,ll'}}= B_1+B_2\frac{\partial\ln F_{ll'}}{\partial n_s}\, .
 \end{eqnarray}
 We then insert this expression into the expectation value of Eq.~\ref{E:estgapp} and, by using $\VEV{g_{2M}}=B_1+B_2\ln(k_{\rm{eff}}/k_*)$, we find
 \begin{eqnarray} \label{E:logk}
 \ln(k_{\rm{eff}}/k_*) = \frac{\sum_l lT_l(\partial\ln C_l/\partial n_s)}{\sum_l lT_l}\, .
 \end{eqnarray}
 We can find the derivative in this expression numerically by using the finite difference method with a two-sided derivative centered at the fiducial value $n_s=1$ with endpoints $n_s^\pm=1\pm0.025$.
 
We use the fiducial cosmological parameters (except for $n_s$) to calculate $C_l$ using {\sc CAMB} \cite{camb}, along with WMAP's instrumental parameters 
$\sigma_T=1.09\times 10^{-5}$ and $\theta_{\rm{fwhm}}=21'$, specifically in the V band.  These parameters give us $T_l$s in the range of $T_2=0.99994$ to 
$T_{400}=0.17$.  We select the pivot wavenumber $k_*=0.002$ Mpc$^{-1}$.  This gives us $k_{\rm{eff}}^{\rm{CMB}}=0.0204$ Mpc$^{-1}$.

Finding $k_{\rm{eff}}$ for the LRG analysis is similar except we include measurements of perturbations at 8 redshift slices instead of only one last-scattering surface.  By making the necessary changes for a galaxy survey analysis, including $F_{g,l(l\pm2)}^i\simeq-C_{g,l}^i$ ($i$ denotes the redshift slice), we find
\begin{eqnarray} \label{E:estgapp2}
\widehat{g}_{2M}\simeq\frac{\sum_{l,i} lT_{g,l}^i[\frac{1}{8\pi}\widehat{g}_{2M,ll}^i+\frac{3}{16\pi}(\widehat{g}_{2M,l(l+2)}^i+\widehat{g}_{2M,l(l-2)}^i)]}{\frac{1}{2\pi}\sum_{l,i} lT_{g,l}^i} \, ,
\end{eqnarray}
where $T_{g,l}^i=[1+C_{g,l}^{i,n}/C_{g,l}^i]^{-2}$, $C_{g,l}^{n,i}=\Delta\Omega/\overline{n}_i$, $\Delta\Omega$ is the pixel size, and $\overline{n}_i$ is the average number of galaxies per pixel.  The expectation value of the estimator $\widehat{g}_{2M,ll'}^i$ is given by
\begin{eqnarray}\label{E:estgl2}
\VEV{\widehat{g}_{2M,ll'}^i} = \frac{\int g_{2M}(k)\bar P_g(k) W_l^i(k)W_{l'}^i(k)k^2dk}{\int \bar P(k) W_l^i(k)W_{l'}^i(k)k^2dk}\, ,
\end{eqnarray}
where $W_l^i(k)$ is the (survey) window function.  $g_{2M}(k)$ again varies smoothly with $\ln k$, which allows us to write
\begin{eqnarray} \label{E:gvar2}
g_{2M}(k) = D_1+D_2\ln k\, .
\end{eqnarray}
This parameterization gives us
\begin{eqnarray} \label{E:estglb2}
\VEV{\widehat{g}_{2M,ll'}^i}= D_1+D_2\frac{\int \ln k\bar P_g(k) W_l^i(k)W_{l'}^i(k)k^2dk}{\int \bar P_g(k) W_l^i(k)W_{l'}^i(k)k^2dk}\, .
\end{eqnarray}
The denominator in the second term is just $F_{g,ll'}^i$, and the numerator can be calculated directly.  Inserting this into the expectation value of Eq.~\ref{E:estgapp2}, which equals $D_1+D_2\ln(k_{\rm{eff}})$, we find $k_{\rm{eff}}^{\rm{LRG}}=0.151$ Mpc$^{-1}$.

\end{document}